\documentclass[10pt]{article}
\usepackage[margin=0.8in]{geometry}
\usepackage{authblk}
\usepackage{sectsty}
\usepackage{amsmath}
\usepackage{graphicx}
\usepackage{enumerate}
\usepackage{natbib}
\usepackage{url}
\usepackage{hyperref}
\usepackage{amsfonts,amssymb}
\usepackage{subfig,etoolbox}
\sectionfont{\large}
\subsectionfont{\large}
\numberwithin{equation}{section}
\newtheorem{def1}{Definition}
\numberwithin{def1}{section}
\newtheorem{theo1}{Theorem}
\numberwithin{theo1}{section}

\title{\fontsize{15}{18}\selectfont \textbf{Skew-elliptical copula based mixed models for non-Gaussian longitudinal data with application to an HIV-AIDS study}}
\author{Subhajit Chattopadhyay}
\affil{\small Bhubandanga, Bidya Sagar Path, Bolpur, West Bengal, India, PIN: 731204}
\begin{document}
\bigskip
\maketitle
\begin{abstract}
This study was sparked by an extensive longitudinal dataset focusing on HIV CD4 T$^+$ cell counts from Livingstone district, Zambia. Analysis of the corresponding histogram plots reveals an absence of symmetry in the marginal distributions, while pairwise scatter plots uncover non-elliptical dependence patterns. Traditional linear mixed models designed for longitudinal data fail to capture these complexities adequately. Therefore, it appears prudent to explore a broader framework for modeling such data. In this article, we delve into generalized linear mixed models (GLMM) for the marginals (e.g., the Gamma mixed model), and we address the temporal dependency of repeated measurements by utilizing copulas associated with skew-elliptical distributions (such as the skew-normal/skew-$t$). Our proposed class of copula-based mixed models simultaneously accommodates asymmetry, between-subject variability, and non-standard temporal dependence, thus offering extensions to the standard linear mixed model based on multivariate normality. We estimate the model parameters using the IFM (inference function of margins) method and outline the process of obtaining standard errors for parameter estimates. Through extensive simulation studies covering skewed and symmetric marginal distributions and various copula choices, we assess the finite sample performance of our approach. Finally, we apply these models to the HIV dataset and present our findings.
\end{abstract}
\section{Introduction}
Biomedical research frequently yields data sets where predictors and response variables are recorded across multiple time points, constituting longitudinal data analysis. Among the various methodologies employed for modeling univariate longitudinal data, linear mixed models (LMMs) stand out as the most prevalent. The seminal work of \citet{laird1982random} introduced the normal linear mixed model, which extends classical linear models by incorporating subject-specific random effects alongside fixed effects. This pioneering approach has since been widely adopted by statisticians across numerous applications, as evidenced by the extensive literature, including works by \citet{verbeke1997linear} and \citet{fitzmaurice2008longitudinal}.
\par Let's consider a scenario with $m$ individuals comprising the subject set. Each individual, indexed by $i = 1,\dots,m$, undergoes repeated measurements and predictors recorded at $n_i$ time points. At each of these time points, the corresponding predictors are presented as a row vector of dimension $1\times p$, forming an $n_i\times p$ matrix denoted as $\mathbf{X}_i$, alongside a $p\times 1$ vector of fixed effects $\mathbf{\beta}$. We denote the $j$-th row of $\mathbf{X}_i$ as $\mathbf{x}_{ij}$. The responses for the $n_i$ time points are arranged as an $n_i \times 1$ column vector, denoted $\mathbf{Y}_i$. Additionally, let $\mathbf{D}_i$ ($n_i\times q$) represent the design matrix corresponding to the random effects, and $\mathbf{b}_i$ of dimension $q$ denote these random effects. Hence, the standard linear mixed model is expressed as -
\begin{equation}\label{lmequ1}
\mathbf{Y}_i = \mathbf{X}_i\mathbf{\beta} + \mathbf{D}_i\mathbf{b}_i + \mathbf{\epsilon}_i, \quad i = 1,\dots,m,
\end{equation}
where $\mathbf{\epsilon}_i$ ($n_i\times q$) signifies the error term. To ensure model identifiability, we assume independence between the collections $\{\mathbf{b}_i: i=1, \ldots ,m\}$ and $\{\mathbf{\epsilon}_i: i=1, \ldots ,m\}$. In the simplest setup, we assume multivariate normality of the random effects and error terms as follows -
\begin{equation}\label{normality}
\mathbf{b}_i \sim N_q(\mathbf{0},\mathbf{\Omega}_b), \quad \mathbf{\epsilon}_i \sim N_{n_i}(\mathbf{0},\mathbf{\Psi}_i),
\end{equation}
where $\mathbf{\Omega}_b$ and $\mathbf{\Psi}_i$ represent the associated dispersion matrices capturing between-subject and within-subject variability. However, in various scenarios, exploratory data analysis (e.g., histograms or pair-wise scatter plots) may suggest that the normality assumptions of this setup might not be appropriate. Generalized linear mixed models allow for different types of response distributions from the exponential family. However, it's worth noting that these models rely on the conditional independence assumption of the response variables given the random effects \citet{mcculloch2003generalized}.
\par Copulas have gained increasing traction in the realm of multivariate analysis, offering a distinct advantage by allowing for separate modeling and estimation of joint dependence structure and marginal distributions. This flexibility has spurred their popularity in recent literature. \citet{lambert2002copula} pioneered the development of a Gaussian copula-based model tailored for multivariate longitudinal data. \citet{masarotto2012gaussian} expanded upon this with Gaussian copula-based regression models designed for non-normally distributed dependent observations. More recently, \citet{killiches2018ad} employed D-vine copulas to model dependence among repeated measurements in unbalanced longitudinal data. Meanwhile, \citet{kurum2018time} proposed a Gaussian copula-based model for mixed longitudinal responses, allowing all model parameters to vary with time. In many biomedical datasets, where pair-wise scatter plots or normal score plots reveal asymmetric shapes, the conventional use of elliptical copulas like Gaussian or Student-$t$ may not be ideal, as they might fail to accurately capture dependency among time points. Consequently, skew-elliptical distributions, as suggested by \citet{azzalini2013skew}, offer an appealing alternative for introducing flexible dependence structures into the model. Additionally, these multivariate copulas can describe various aspects of the model's dependence structure, including reflection, permutation asymmetry, and tail dependence, if present in the data, as highlighted by \citet{chang2020copula}. This underscores the versatility and applicability of copula-based approaches in capturing complex dependence structures across various types of longitudinal data.
\par Motivated by recent HIV-AIDS CD4 count data from Livingstone district, Zambia, this study introduces a copula-based longitudinal model that incorporates temporal dependence through skew-elliptical copulas. We extend the multivariate normal linear mixed model outlined in Equation \ref{lmequ1} by allowing generalized linear mixed models (GLMMs) as marginals, while characterizing the dependence structure using skew-elliptical distributions. The remainder of the paper is structured as follows. Section \ref{sec2} offers an in-depth exploration of the dataset under analysis, providing crucial details for our study. Moving to Section \ref{sec3}, we delineate our innovative modeling framework, which integrates GLMMs as marginals and leverages skew-elliptical distributions to capture the intricate temporal dependencies. In Section \ref{sec4}, we present a comprehensive overview of skew-elliptical distributions and their associated copulas, setting the stage for our modeling approach. In Section \ref{sec5}, we delve into the estimation of model parameters using the inference function of margins (IFM) method and discuss the corresponding asymptotic normality. Section \ref{sec7} presents simulation studies aimed at assessing parameter estimation performance for the proposed class of models, considering both skewed and symmetric marginal mixed models across different sample sizes. Finally, in Section \ref{sec8}, we draw conclusions from our findings and engage in a broader discussion, highlighting the implications and potential avenues for future research.
\section{HIV CD4 positive T cell count data}\label{sec2}
Human Immunodeficiency Virus (HIV) is a viral infection that progressively weakens the immune system, leading to Acquired Immunodeficiency Syndrome (AIDS). Regrettably, despite extensive research efforts, a clinically proven vaccine for this virus remains elusive to date. Consequently, individuals rely on available antiviral medications to impede viral replication. Among the pivotal markers used to assess the efficacy of antiviral therapies are HIV-1 RNA copies and CD4 T$^+$ cell counts. Given the skewed distribution of these markers, researchers often opt to model them using skew-elliptical distributions. \citet{lin2013multivariate} utilized a multivariate skew-normal mixed model in the ACTG $315$ study to effectively model these markers. Similarly, \citet{bandyopadhyay2012skew} explored a skew-normal based linear mixed model in a study focusing on HIV viral load. These approaches highlight the importance of leveraging skew-elliptical distributions in modeling HIV-related data, offering valuable insights into the complex dynamics of the disease and its treatment.
\par
\begin{figure}[h]
    \centering
    \includegraphics[width = 7cm]{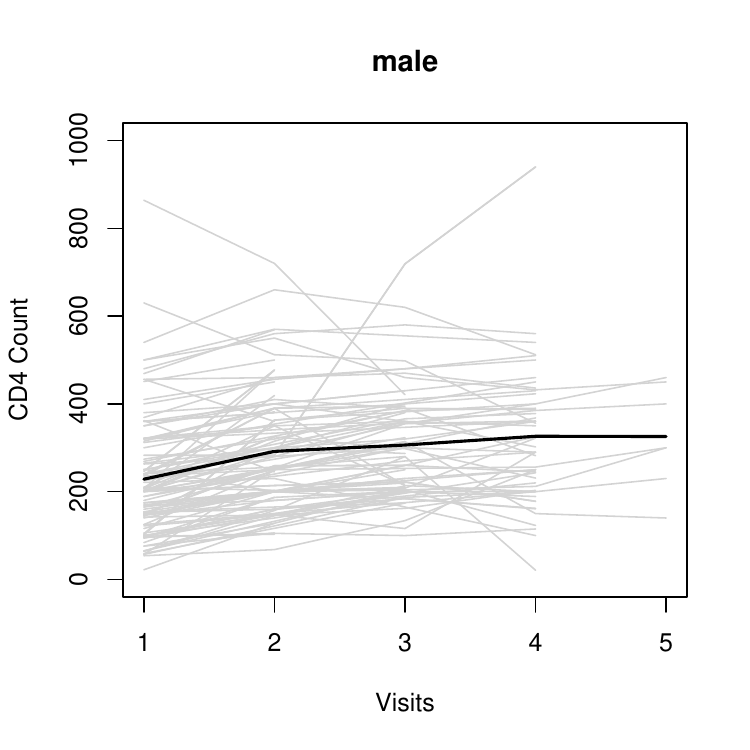}
    \includegraphics[width = 7cm]{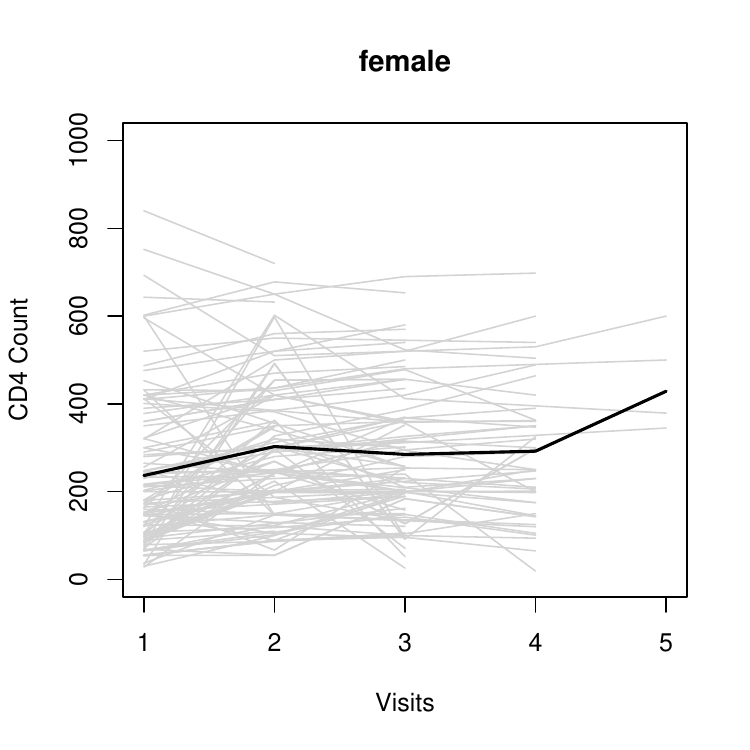}
    \caption{Individual and average profiles for male (left panel) and female (right panel) patients over time starting from initial. Black lines represent the mean profiles.}
    \label{fig:profiteplot1}
\end{figure}

\begin{figure}[t]
    \centering
    \includegraphics[width = 14cm]{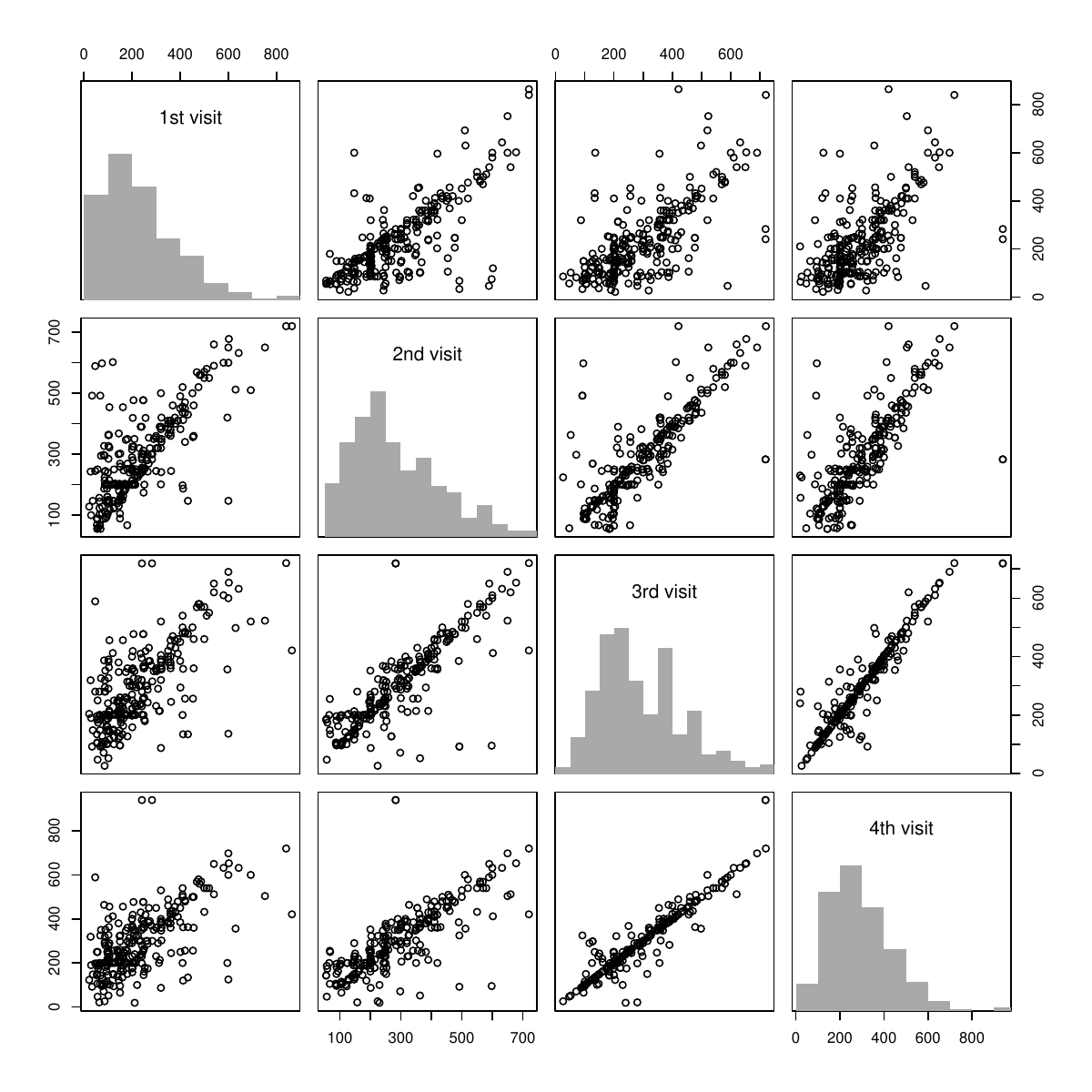}
    \caption{Pairwise scatter plots of the CD4 counts of the patients for first 4 visits.}
    \label{fig:histpairs1}
\end{figure}

 The motivating dataset analyzed in this article was sourced from \href{https://data.mendeley.com/datasets/f7wfdbrfys/1}{Mendeley}, focusing on a recent HIV-CD4 study from the Livingstone district, Zambia ($2016$). These data were collected during a survey of antiretroviral (ARV) combination treatments for HIV, as part of the PhD thesis work of Urban N. Haankuku at the University of South Africa. They have delved into the performance of ARV combinations on HIV-naive patients using various models. According to the World Health Organization (WHO), HIV-AIDS remains a leading cause of death in Zambia, with approximately a million deaths attributed to HIV-AIDS-related causes. Left untreated, the disease can lead to a reduction in CD4 T$^+$ cell clusters and an increase in HIV viral load. With no permanent cure available to date, the primary option is to administer antiretroviral drugs to mitigate immune suppression. The dataset comprises CD4 counts of $261$ HIV-naive patients measured every twelve weeks from the initial diagnosis over a span of $48$ weeks. These patients were administered three different ARV combinations as part of their first baseline regimen (FBR). Covariates such as gender, age, and initial weight of each patient were also recorded. Figure \ref{fig:profiteplot1} illustrates the evolution of CD4$^+$ T cell counts over time. While the mean structure appears nearly linear, significant variability is evident in the between-subject responses. In Figure \ref{fig:histpairs1}, histograms of each time point are plotted on the diagonals, with pair-wise scatter plots depicted on the off diagonals. Notably, based on the initial diagnosis, the marginals exhibit skewness with positive real support, and the scatter plots reveal non-elliptical dependence patterns. This is further confirmed by the normal scores, indicating reflection and permutation asymmetry, along with stronger dependence than Gaussian in the joint upper and lower tails. Given the skewed nature of the CD4 counts marker with positive real support, we opt for a Gamma mixed model for the marginals. For comparison, we also employ a normal mixed model as the marginals.
\section{A copula based longitudinal model}\label{sec3}
Let the response variables $\mathbf{Y}_i = (Y_{i1},\dots,Y_{in_i})^\intercal, i = 1,\dots,m,$ follow an $n_i$-variate distribution with predefined mean and dispersion matrix. Suppose observations from different individuals are independent, and to account for subject's individual effects we consider the distribution of $\mathbf{Y}_i$ conditional on $\mathbf{b}_i$ as
\begin{equation}\label{model1}
\mathbf{Y}_i|\mathbf{b}_i \sim F_{n_i}(\eta(\mathbf{X}_i\mathbf{\beta} + \mathbf{D}_i\mathbf{b}_i),\mathbf{\Sigma}(\xi_i,\mathbf{t}_i))
\end{equation}
where $\mathbf{\xi}_i$ is the auto-regressive time variant parameter with respect to the time points, $\mathbf{t}_i = (t_{i1},\dots,t_{in_i})^\intercal$ and $\eta(.)$ is a known link function. $F_n(\eta,\mathbf{\Sigma})$ is an $n$-variate distribution function with mean $\eta$ and covariance $\mathbf{\Sigma}$. Furthermore, $\mathbf{X}_i:n_i \times p$ and $\mathbf{D}_i:n_i \times q$ are the known design matrices as described earlier. We assume the marginal densities of $Y_{ij}|\mathbf{b}_i$ are from the exponential family and functions of $\{\mathbf{x}_{ij},\mathbf{\beta},t_{ij},\mathbf{d}_{ij},\mathbf{b}_i\}$ via the same known link $\mathbf{\eta}(.)$. In this article we assume the random effects are independent and normally distributed, i.e. $\mathbf{b}_i \sim N_q(0,\mathbf{\Omega}_b)$. We attempt to model such distribution using a copula based GLMM as
\begin{equation}\label{copglmm1}
F_{n_i}(\mathbf{y}_i|\mathbf{b}_i,\mathbf{\theta}^*_i) = C_{n_i}(F(y_{i1}|\mathbf{b}_i,\theta_{i1}),\dots,F(y_{in_i}|\mathbf{b}_i,\theta_{in_i})|\mathbf{\phi}_i)     
\end{equation}
where $\mathbf{\theta}^*_i = (\mathbf{\theta}_i^\intercal,\mathbf{\phi}_i^\intercal)^\intercal$ is the set of all vector valued parameters in the conditional model. $\mathbf{\theta}_i$ accounts for all the parameters present in the marginals and $\mathbf{\phi}_i$ accounts for the dependence parameters. Then the corresponding density function is given by
\begin{equation}\label{copden1}
f_{n_i}(\mathbf{y}_i|\mathbf{b}_i,\mathbf{\theta}^*_i) = c_{n_i}(F(y_{i1}|\mathbf{b}_i,\theta_{i1}),\dots,F(y_{in_i}|\mathbf{b}_i,\theta_{in_i})|\mathbf{\phi}_i)\prod_{j=1}^{n_i}f(y_{ij}|\mathbf{b}_i,\theta_{ij}).
\end{equation}
\par Copula-based models offer remarkable flexibility in analyzing the temporal dependence of longitudinal data, as the responses at each time point are associated with a predefined marginal distribution. Essentially, a copula can be conceptualized as an association function that characterizes the dependence between separately specified marginals. With fixed marginals, various multivariate models can be derived by considering different copula functions. The random effects $\mathbf{b}_i$ are interpreted as the unobserved regression parameters for the $i$-th subject, which account for variability between subjects. The subsequent section outlines a class of multivariate copulas possessing desirable dependence properties, which can be effectively utilized for the models outlined in Equation \ref{copglmm1}.
\section{Skew-elliptical distributions and related copulas}\label{sec4}
Multivariate skew-normal and related family of distributions proposed by \citet{azzalini2013skew} have been applied in several bio-medical studies to model non-Gaussian data. Copulas generated from skew-elliptical class of distributions can provide various flexible dependence structures. These multivariate copulas are reflection as well as permutation asymmetric and can be used to describe a general dependence structure of the model.
\begin{def1}
A random variable $\mathbf{Z} \in \mathcal{R}^d$ is said to have a mean zero skew-normal distribution, denoted as $\mathbf{Z} \sim SN_d(0,\mathbf{\Sigma,\lambda})$, if is continuous with the probability density function,
\begin{equation}\label{sndef1}
sn_d(\mathbf{z|\Sigma,\lambda}) = 2\phi_d(\mathbf{z|\Sigma})\Phi_1\Big(\mathbf{\lambda}^\intercal\mathbf{\Sigma}^{-1/2}\mathbf{z}\Big)
\end{equation}
for $\mathbf{\lambda} \in \mathcal{R}^d$ and $\mathbf{\Sigma}$ be a $d\times d$ positive definite matrix. For the univariate case
\begin{equation}\label{sndef2}
sn_1(z|\sigma^2,\lambda) = 2\phi_1(z|\sigma^2)\Phi_1\big(\frac{\lambda z}{\sigma}\big)
\end{equation}
where $\phi_d(.)$ and $\Phi_d(.)$ are the PDF and CDF of a $d$-dimensional standard multivariate normal variable, respectively.
\end{def1}
Skew-normal copula is obtained from the multivariate skew-normal distribution and the corresponding univariate quantiles using the probability integral transformation method. \citet{wei2016multivariate} studied a class of copulas generated from skew-normal distribution.
\begin{def1}\label{sncopdef1}
A $d$-dimensional copula is said to be a skew-normal copula if
\begin{equation}\label{csndef1}
C_{d,SN}(\mathbf{u}|\mathbf{\Sigma,\lambda}) = SN_d(SN_1^{-1}(u_1|1,\lambda^*_1),\dots,SN_1^{-1}(u_d|1,\lambda^*_d)|\mathbf{\Sigma,\lambda})
\end{equation}
where $SN_1^{-1}(u_j|1,\lambda^*_j)$ denotes the inverse of the CDF of $Z_j \sim SN_1(1,\lambda^*_j)$ distribution for $j = 1,\dots,d$. The corresponding skew-normal copula density is given by
\begin{equation}\label{csndef2}
c_{d,SN}(\mathbf{u}|\mathbf{\Sigma,\lambda}) = \frac{sn_d(SN_1^{-1}(u_1|1,\lambda^*_1),\dots,SN_1^{-1}(u_d|1,\lambda^*_d)|\mathbf{\Sigma,\lambda})}{\prod_{j=1}^d sn_1(SN_1^{-1}(u_j|1,\lambda^*_j))}.
\end{equation}
\end{def1}
Here the skewness parameters ($\lambda^*_j$) of the univariate quantiles can be obtained from the multivariate parameters $\mathbf{\Sigma,\lambda}$ by
\begin{equation}\label{trnsf1}
\lambda^*_j = \frac{\delta^*_j}{\sqrt{1 - \delta^{*2}_j}}, \;\; \text{where} \;\; \mathbf{\delta}^* = \mathbf{\Sigma}^{1/2}\frac{\mathbf{\lambda}}{\sqrt{1 + \mathbf{\lambda}^\intercal\mathbf{\lambda}}}.
\end{equation}
For more details regarding this, one can go through \citet{azzalini2003distributions}. Skew-normal copula is exchangeable or permutation symmetric if and only if $\lambda_j = \lambda$ for all $j = 1,\dots,d$, and all off-diagonal elements of the correlation matrix $\mathbf{\Sigma}$ are equal. Note that Gaussian copula is nested to the skew-normal copula when $\lambda_j = 0$ for all $j = 1,\dots,d$.
\par Multivariate skew-$t$ distribution is a member of the skew-elliptical family of distribution which is defined as a scale mixture of skew-normal distribution. \citet{gupta2003multivariate} and \citet{sahu2003new} discussed the theoretical properties of this distribution.
\begin{def1}
Let $\mathbf{Z} \in \mathcal{R}^d$ be a mean zero skew-normal variable and $\mathbf{V}$ be another variable independent with $\mathbf{Z}$ such that, $\mathbf{V} \sim \mathbf{\chi}^2_\mathbf{\nu} /\mathbf{\nu}$. Then $\mathbf{T} = \mathbf{V}^{-1/2}\mathbf{Z}$ follows a mean zero skew-$t$ distribution with the probability density function,
\begin{equation}\label{stdef1}
st_d(\mathbf{t}|\mathbf{\Sigma,\lambda,\nu}) = 2t_d(\mathbf{t}|\mathbf{\Sigma,\nu})T_1\Bigg(\mathbf{\lambda}^\intercal\mathbf{\Sigma}^{-1/2}\mathbf{t}\sqrt{\frac{\mathbf{\nu} + d}{\mathbf{Q_t} + \mathbf{\nu}}}\Big| \mathbf{\nu} + d\Bigg) 
\end{equation}
where $\mathbf{Q_t} = \mathbf{t}^\intercal\mathbf{\Sigma}^{-1}\mathbf{t}$, $\mathbf{\lambda} \in \mathcal{R}^d$ and $\mathbf{\Sigma}$ be a $d\times d$ positive definite matrix. For the univariate case
\begin{equation}
st_1(t|\sigma^2,\lambda,\nu) = 2t_1(t|\sigma^2,\nu)T_1\Big(\frac{\lambda t}{\sigma}\sqrt{\frac{\nu + 1}{Q_t + \nu}}\Big| \nu + 1\Big)
\end{equation}
where $t_d(.)$ and $T_d(.)$ are the PDF and CDF of a $d$-dimensional standard Student-$t$ variable, respectively.
\end{def1}
Multivariate skew-$t$ copula can be similarly obtained using the above distribution as -
\begin{def1}\label{stcopdef1}
A $d$-dimensional copula is said to be a skew-$t$ copula if
\begin{equation}\label{cstdef1}
C_{d,ST}(\mathbf{u}|\mathbf{\Sigma,\lambda,\nu}) = ST_d(ST_1^{-1}(u_1|1,\lambda^*_1,\nu),\dots,ST_1^{-1}(u_d|1,\lambda^*_d,\nu)|\mathbf{\Sigma,\lambda,\nu})
\end{equation}
where $ST_1^{-1}(u_j|1,\lambda^*_j,\nu)$ denotes the inverse of the CDF of $T_j \sim ST_1(1,\lambda^*_j,\nu)$ distribution. The corresponding skew-t copula density is given by
\begin{equation}\label{cstdef2}
c_{d,ST}(\mathbf{u}|\mathbf{\Sigma,\lambda,\nu}) = \frac{st_d(ST_1^{-1}(u_1|1,\lambda^*_1,\nu),\dots,ST_1^{-1}(u_d|1,\lambda^*_d,\nu)|\mathbf{\Sigma,\lambda,\nu})}{\prod_{j=1}^d st_1(ST_1^{-1}(u_j|1,\lambda^*_j,\nu))}.
\end{equation}
\end{def1}
\par
\begin{figure}[h]
    \centering
    \includegraphics[width = 14cm]{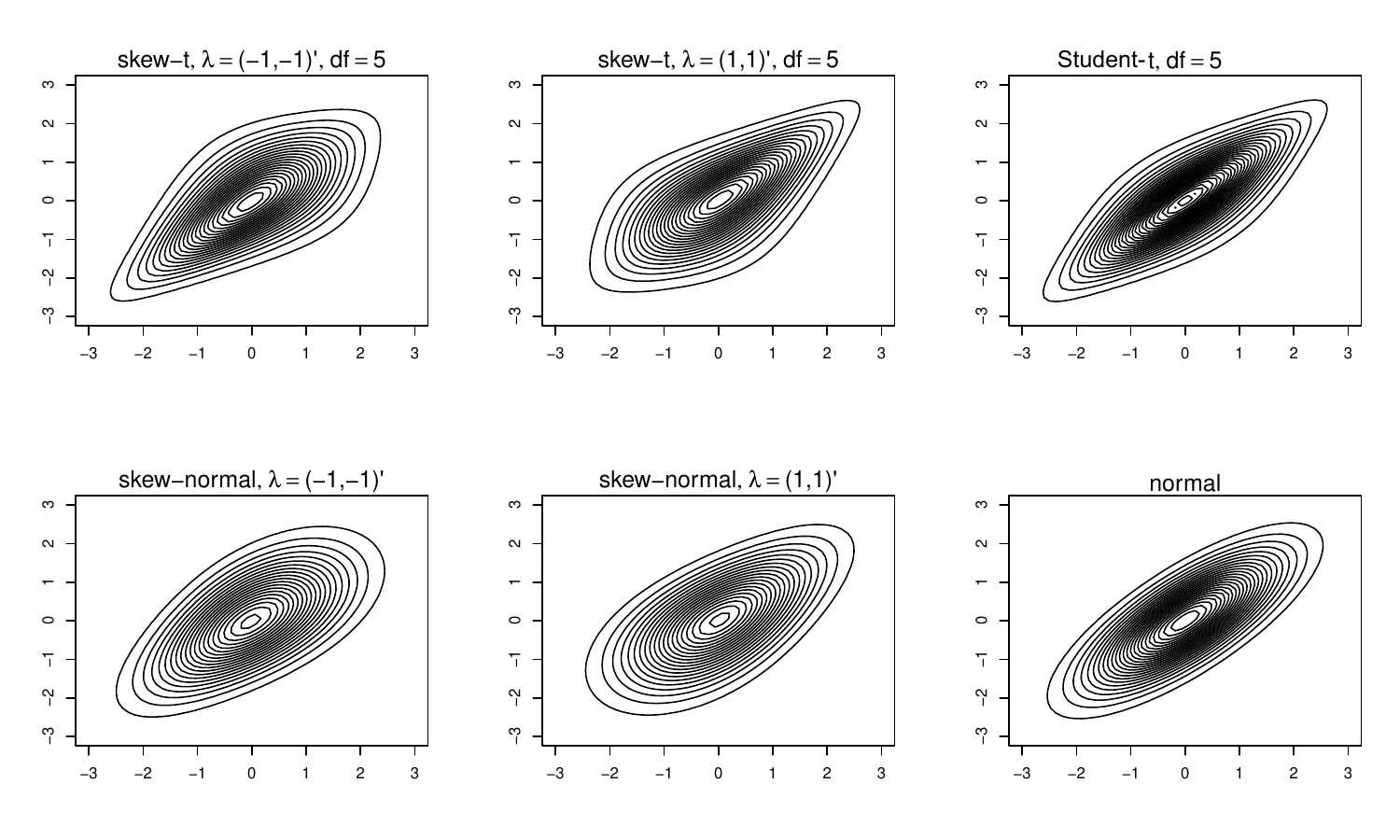}
    \caption{Contour plots of bivariate distributions using skew-$t$ copula (on the upper row) (with $\nu = 5$ and $\lambda = \{(-1,-1),(1,1),(0,0)\}$) with corresponding skew-normal copula (on the lower row); the common correlation parameter $\rho = 0.77$ and common marginals are standard normal.}
    \label{fig:copcontour1}
\end{figure}

 The skew-normal copula, as defined in Definition (\ref{sncopdef1}), encapsulates the non-exchangeable dependence between the variables of interest. Here, the correlation matrix $\mathbf{\Sigma}$ captures the association between unobservable or latent variables $Z_j$ in (\ref{csndef1}), while $\mathbf{\lambda} = (\lambda_1,\dots,\lambda_d)^\intercal$ accounts for the differential skewness of the involved variables. In contrast, the skew-$t$ copula, as described in Definition (\ref{stcopdef1}), introduces an additional parameter known as the "degrees of freedom," which accommodates possible tail dependence in the data. \citet{yoshiba2018maximum} delved into the applications and maximum likelihood estimation of the skew-$t$ copula, while \citet{smith2012modelling} discussed its Bayesian estimation. Notably, as $\mathbf{\nu} \to \infty$, the skew-normal copula is obtained. Moreover, the Gaussian and Student-$t$ copulas are nested within the skew-$t$ copula. To provide visual insights, we depict the contours of the joint densities of the skew-$t$, Student-$t$, skew-normal, and normal copulas, utilizing standard normal margins, as shown in Figure \ref{fig:copcontour1}.
\section{Parameter estimation}\label{sec5}
In Section \ref{sec3}, we outlined a class of copula-based longitudinal models. However, parameter estimation for such models, which involve complex non-exchangeable multivariate copulas as described in Section \ref{sec4}, is often challenging, especially in moderate to high-dimensional scenarios. Direct maximum likelihood estimation is not always feasible due to the computationally intensive nature of computing the quantiles of skew-elliptical copulas. Hence, we resort to a fully parametric estimation method known as Inference Function for Margins (IFM) to estimate the parameters for the models outlined in Equation (\ref{copglmm1}). When the likelihood function of the data analytic model poses computational challenges, \citet{joe1996estimation} introduced the two-stage maximum likelihood estimation method. In this approach, all parameters present in the model are estimated in two steps. \citet{joe2005asymptotic} further discussed the asymptotic efficiency of this method, wherein the univariate parameters are estimated from separate univariate likelihoods at the first stage. Subsequently, the multivariate parameters are estimated from the multivariate likelihood with the univariate parameters given the values from the first stage.
\par For the class of models outlined in (\ref{copglmm1}), we employ the IFM method to estimate the model parameters and their standard errors using the corresponding asymptotic covariance matrix. Given that the response variables $\mathbf{Y}_i$ are conditionally independent, the joint density of the $i$-th response is represented as 
\begin{equation}
f_{n_i}(\mathbf{y}_i,\mathbf{b}_i|\mathbf{\theta}^*_i) = f_{n_i}(\mathbf{y}_i|\mathbf{b}_i,\mathbf{\theta}^*_i) g(\mathbf{b}_i).
\end{equation}
In the presence of unobserved random effects, in the first stage we consider each marginal responses $y_{ij}|\mathbf{b}_i$ are conditionally independent. Therefore, we construct the inference functions as
\begin{align}\label{imffun1}
l^*_i(\mathbf{\theta}_i|\mathbf{y}_i) & = \log\int \prod_{j=1}^{n_i} f(y_{ij}|\mathbf{b}_i,\theta_{ij}) g(\mathbf{b}_i) d\mathbf{b}_i, \nonumber \\ l^*_i(\mathbf{\theta}^*_i|\mathbf{y}_i) & = \log\int c_{n_i}(F(y_{i1}|\mathbf{b}_i,\theta_{i1}),\dots,F(y_{in_i}|\mathbf{b}_i,\theta_{in_i})|\mathbf{\phi}_i) \prod_{j=1}^{n_i} f(y_{ij}|\mathbf{b}_i,\theta_{ij}) g(\mathbf{b}_i) d\mathbf{b}_i.
\end{align}
Now to obtain the IFM estimates based on $m$ independent observations we assume, $\mathbf{\theta}^*_i = (\mathbf{\theta}_i^\intercal,\mathbf{\phi}_i^\intercal)^\intercal$ are all functions of $\mathbf{\theta}^* = (\mathbf{\theta}^\intercal,\mathbf{\phi}^\intercal)^\intercal$ for $i = 1,\dots,m$. For notational simplicity, we assume that the parameters of the random effects distribution is included in $\mathbf{\theta}$ and $\mathbf{\phi}$ includes the dependence parameters as in (\ref{copden1}). The IFM method estimates the model parameters by
\begin{align}\label{ifm1}
\hat{\mathbf{\theta}} & = \arg\underset{\mathbf{\theta}}{\max} \sum_{i=1}^m l^*_i(\mathbf{\theta}_i|\mathbf{y}_i), \nonumber \\ \hat{\mathbf{\phi}} & = \arg\underset{\mathbf{\phi}}{\max} \sum_{i=1}^m l^*_i(\hat{\mathbf{\theta}_i},\mathbf{\phi}_i|\mathbf{y}_i)
\end{align}
and the estimating equations for $\mathbf{\theta}$ and $\mathbf{\phi}$ based on IFM are
\begin{align}\label{estequ1}
\Psi_{1m}(\mathbf{\theta}) = \sum_{i=1}^m \Psi_{i1m}(\mathbf{\theta}) & = \sum_{i=1}^m \frac{\partial}{\partial\mathbf{\theta}} l^*_i(\mathbf{\theta}_i|\mathbf{y}_i) = \mathbf{0}, \nonumber \\ \Psi_{2m}(\mathbf{\phi}) = \sum_{i=1}^m \Psi_{i2m}(\mathbf{\phi}) & = \sum_{i=1}^m \frac{\partial}{\partial\mathbf{\phi}} l^*_i(\hat{\mathbf{\theta}_i},\mathbf{\phi}_i|\mathbf{y}_i) = \mathbf{0}
\end{align}
provided the derivatives exist. The IFM estimates are obtained either by numerically maximizing (\ref{ifm1}) or by solving the system of non-linear equations as in (\ref{estequ1}). 
\section{Asymptotic normality}
Here we show under some regularity conditions, the IFM estimators for the class of models in (\ref{copglmm1}) are consistent and asymptotically normal. We also present the theoretical analysis for independent and non-identically distributed (i.n.i.d) observations including random effects. Let us denote the true value of the parameters $\mathbf{\theta^*} = (\mathbf{\theta}^\intercal,\mathbf{\phi}^\intercal)^\intercal$ by $\mathbf{\theta}^*_0 = (\mathbf{\theta}_0^\intercal,\mathbf{\phi}_0^\intercal)^\intercal$ and let $\Psi^*_{im} = (\Psi_{i1m}^\intercal,\Psi_{i2m}^\intercal)^\intercal$ be the stack of the vector valued inference function for the $i$-th response. To establish the consistency and asymptotic normality, we need the set of assumptions in Appendix \ref{apndx1}, which make $\Psi^*_m = \sum_{i=1}^m \Psi^*_{im}$ a regular inference function vector and consequently we have the following theorem.
\begin{theo1}\label{theorem1}
Consider the model (\ref{copglmm1}) and let $\hat{\mathbf{\theta}}^* = (\hat{\mathbf{\theta}}^\intercal,\hat{\mathbf{\phi}}^\intercal)^\intercal$ denote the IFME of $\mathbf{\theta^*}$ corresponding to IFM (\ref{imffun1}). Under the assumptions in Appendix \ref{apndx1}, $\hat{\mathbf{\theta}}^*$ is a consistent estimator of $\mathbf{\theta^*}$. Furthermore, as $m \rightarrow \infty$, we have asymptotic normality as
\begin{equation}\label{assnorm1}
\sqrt{m}(\hat{\mathbf{\theta}}^* - \mathbf{\theta}^*_0) \;\; \overset{d}{\rightarrow} N(\mathbf{0},J_\Psi(\mathbf{\theta}^*_0)^{-1}), \;\; \text{where} \;\; J_\Psi(\mathbf{\theta}^*_0) = D_\Psi(\mathbf{\theta}^*_0)^\intercal M_\Psi(\mathbf{\theta}^*_0)^{-1} D_\Psi(\mathbf{\theta}^*_0), \nonumber
\end{equation}
\begin{equation}
M_\Psi(\mathbf{\theta}^*_0) = \lim_{m \rightarrow \infty} \frac{1}{m}\sum_{i=1}^m E[\Psi_{im}(\mathbf{\theta}^*_0)\Psi_{im}(\mathbf{\theta}^*_0)^\intercal] \;\; \text{and} \;\; D_\Psi(\mathbf{\theta}^*_0) = \lim_{m \rightarrow \infty} \frac{1}{m}\sum_{i=1}^m E\Big[\frac{\partial}{\partial\mathbf{\theta}^*}\Psi_{im}(\mathbf{\theta^*})\Big|_{\mathbf{\theta}^*_0}\Big]. \nonumber
\end{equation}
\end{theo1}
Based on the general discussions in \citet{xu1996statistical}, we also provide a simplified proof of the above theorem in Appendix \ref{apndx1} Note that, here $M_\Psi(\mathbf{\theta}^*_0)$ and $D_\Psi(\mathbf{\theta}^*_0)$ both depends on the distribution of the random effects $\mathbf{b}_i$. The asymptotic covariance matrix in (\ref{assnorm1}) is known as Godambe information matrix in the literature. The observed version of this matrix can be numerically obtained by
\begin{align}\label{obsgod1}
M_\Psi(\hat{\mathbf{\theta}}^*) & = \frac{1}{m}\sum_{i=1}^m \Psi_{im}(\mathbf{\theta^*})\Big|_{\hat{\mathbf{\theta}}^*} \Psi_{im}(\mathbf{\theta^*})^\intercal\Big|_{\hat{\mathbf{\theta}}^*}, \nonumber \\
D_\Psi(\hat{\mathbf{\theta}}^*) & = \frac{1}{m}\sum_{i=1}^m \left(\begin{array}{cc} \frac{\partial}{\partial\mathbf{\theta}} \Psi_{i1m}(\mathbf{\theta})\Big|_{\hat{\mathbf{\theta}}} & \mathbf{0} \\ \mathbf{0} & \frac{\partial}{\partial\mathbf{\phi}}\Psi_{i2m}(\mathbf{\phi})\Big|_{\hat{\mathbf{\phi}}} \end{array}\right),
\end{align}
which leads to a straightforward calculation of the standard errors for the parameter estimates based on IFM (\ref{imffun1}), using the square-roots of the diagonals of $J_\Psi(\hat{\mathbf{\theta}}^*)^{-1}$. We employed numerical derivative methods ({\em numderiv} in R) to obtain the matrices in (\ref{obsgod1}).
\section{Numerical implementation}
As we described in Section \ref{sec3}, we consider generalized linear mixed models for the marginals of the class of models in (\ref{copglmm1}). Gamma distribution with $\log$ link is a good candidate, when the response distributions are seen to be asymmetric with positive real support. The Gamma distributed response variable $Y_{ij}$ with shape parameter $\kappa$ and mean $\eta_{ij}$ has the density
\begin{equation}
f(y_{ij}|\eta_{ij},\kappa) = \Gamma^{-1}(\kappa)\Big(\frac{\eta_{ij}}{\kappa}\Big)^{-1}y_{ij}^{-k}\exp\Big(-\frac{y_{ij}\kappa}{\eta_{ij}}\Big).
\end{equation}
For comparison, we also consider a symmetric marginal distribution such as normal with mean $\eta_{ij}$ and variance $\sigma^2$, having the density
\begin{equation}
f(y_{ij}|\eta_{ij},\sigma^2) = \frac{1}{\sqrt{2\pi}\sigma}\exp\Big(-\frac{1}{2}\Big(\frac{y_{ij}-\eta_{ij}}{\sigma}\Big)^2\Big).
\end{equation}
To evaluate the integrations in (\ref{imffun1}), we use standard Gauss-Hermite quadrature rule with $15$ quadrature points. The matrices in (\ref{obsgod1}) are obtained similarly, to compute the standard errors of the parameter estimates.
\par The serial correlation of the repeated measurements is characterized by the correlation matrix of the skew-elliptical copulas. In Equation (\ref{model1}), $\mathbf{\Sigma}$ is treated as a function of time and the dispersion parameter $\mathbf{\xi}$, representing time-varying serial dependence. For the estimating model described in (\ref{copglmm1}), we assume homogeneous variance ($\sigma^2$ or $1/\kappa$) within units and consider an $AR(1)$ structure for the correlation matrix $\mathbf{\Sigma}$ in the skew-elliptical copulas. This correlation structure is typically utilized when the observations are equally spaced. To construct the correlation matrix $\mathbf{\Sigma}(\xi,\mathbf{t}_i)$ of the skew-elliptical copulas, we implement the exponential autocorrelation function defined as
\begin{equation}\label{autocor1}
Corr(Y_{ij},Y_{ik}) = \exp \left(-\mathbf{\xi}|t_{ij} - t_{ik}|\right), \quad \mathbf{\xi} \geq 0,
\end{equation}
where $\mathbf{\xi}$ denotes the dispersion parameter, and $t_{ij}$ and $t_{ik}$ represent the time points of the $j$-th and $k$-th measurements for the $i$-th subject, respectively. This structural assumption is made based on the sample correlation matrix of our data. Similarly one can adopt an exchangeable correlation matrix depending upon the data. The IFM method estimates model parameters in two stages but is not as efficient as direct maximum likelihood estimation. As highlighted by \citet{kim2007comparison}, even a slight bias from the first stage of IFM estimation can lead to significant disturbances in the copula parameter estimation during the second stage. In our scenario, we observed that leaving the skewness parameter of the skew-normal and skew-$t$ copula unrestricted resulted in either heavily biased estimation of that parameter or failure to find the optimal solution. To address this, we assume the same value for the skewness parameter $\mathbf{\lambda}$ across each dimension (equi-skewness), given that longitudinal data involve the same sample observed across time. These structured assumptions not only reduce the number of estimable parameters in the model but also ensure the positive definiteness of the correlation matrices during the estimation process. For numerical optimization, we utilize the {\em optim} function in R with the L-BFGS-B method. In the subsequent section, we delve into the comparison between different fitted models to the data, elucidating their respective strengths and weaknesses.
\section{Model comparison}
An intrinsic challenge in the analysis of longitudinal data is model selection, particularly in determining the most suitable number of components for a given dataset. For the class of models described in (\ref{copglmm1}), one must choose appropriate marginal distributions in conjunction with the multivariate copula. To aid in this decision-making process, log-likelihood-based measures such as AIC (Akaike Information Criterion) and BIC (Bayesian Information Criterion) are frequently employed. These criteria penalize models with a large number of parameters, thus favoring simpler models. AIC and BIC are defined by -
\begin{equation}\label{crtr1}
AIC = -2l(\hat{\mathbf{\theta}}^*) + 2\dim(\hat{\mathbf{\theta}}^*), \;\; BIC = -2l(\hat{\mathbf{\theta}}^*) + \log(m)\dim(\hat{\mathbf{\theta}}^*)
\end{equation}
where $\hat{\mathbf{\theta}}^*$ represents the maximum likelihood estimates of the model parameters, and $m$ is the sample size. While the penalty of AIC depends solely on the number of parameters in the model, BIC's penalty also accounts for the sample size. However, when employing two-stage maximum likelihood estimation for the class of models, these information criteria are modified accordingly in the literature (\citet{jordanger2014model}). It is assumed that the true data-generating model is included in the set of candidate models. \citet{ko2019copula} demonstrated that if the marginals and the copula family are correctly specified, then the two-stage AIC coincides with the original AIC. Similar justifications are discussed in \citet{killiches2018ad} (Remark 1). In our approach, we use these modified versions of AIC and BIC as close approximations of the true AIC and BIC by substituting the two-stage estimates into the actual likelihood functions evaluated by numerical quadrature integration. This allows us to effectively assess the fit of various candidate models to the data.
\section{Simulation design and analysis}\label{sec7}
Simulation studies are conducted to illustrate the performance of IFM estimation for the proposed class of multivariate models in (\ref{copglmm1}). The goal of this study is to monitor the parameter inference using IFM estimation with different combinations of copulas and marginal distributions. We have used two different sample sizes, $m = \{200,500\}$ and fixed the dimension of each response vectors to $n_i = 4$. Since the true parameters are known, to generate the response variable $\mathbf{Y}_i|\mathbf{b}_i$, we first sample from the multivariate copulas discussed in section \ref{sec4} with similar values of the dependence parameters, $(\mathbf{\xi,\lambda,\nu})^\intercal$. Then we use probability integral transformation to generate per-unit multivariate response $\mathbf{Y}_i|\mathbf{b}_i$. We assume the distribution of the random effects to be normal for all the cases. Mainly due to increase in computation time, we restrict with random intercept structure in the linear predictor of the models. Note that, in (\ref{imffun1}), we use a straight forward GLMM estimation at the first stage, which makes the considered models comparable. While keeping the marginal distributions fixed, we compare the parameter estimates with $4$ choices of the multivariate copulas. To obtain the initial estimates of the marginal parameters we use {\em nlme} package in R, and then the initial values of the dependence parameters are obtained from fitting the rescaled empirical CDF (uniform data) to the copulas. Furthermore, for the skew-$t$ and Student-$t$ copulas we used fixed integer valued degrees of freedom parameter $\mathbf{\nu}$.
\par The following class of models investigates the parameter estimation under both binary and categorical variables present as covariates, in order to mimic the real data discussed in the following section. We denote the general structure of the models with both fixed and random effects with time-varying dependence as
\begin{align}\label{simmodel1}
\mathbf{Y}_i|\mathbf{b}_i & \sim F_{n_i}(\eta(\mathbf{X}_i\mathbf{\beta} + \mathbf{D}_i\mathbf{b}_i),\mathbf{\Sigma}(\xi,\mathbf{t}_i)) \nonumber \\ \text{with} \;\; x_{ij}\mathbf{\beta} + d_{ij}\mathbf{b}_i & = \beta_0 + x_{1i}\beta_1 + x_{2i}\beta_2 + t_{ij}\beta_3 + b_i,
\end{align}
where $F_{n_i}$ is a multivariate distribution with link function $\eta$ as described in (\ref{model1}). We set here $\beta_0 = 1.5$, $\beta_1 = 0.5$, $\beta_2 = 0.5$ and $\beta_3 = 1.0$ for the fixed effect parameters. Here $x_{1i} = 1*I(i\leq m/2) + 2*I(i> m/2)$, and $x_{2i} = 0$ or $1$ assigned randomly ($m$ is the sample size). The random effects are set as $b_i \sim N(0,1)$ and the time points $t_{ij} = j - 2.5$ for $j = 1,\dots,n_i$. In one scenario we use Gamma distribution with log-link and shape parameter $\kappa = 3$ for the marginals. In the other scenario we use normal distribution with standard deviation $\sigma = 1$. For the correlation matrix in all the multivariate copulas we set the time difference per observation within each response vector to unity. Hence the off-diagonal entries of the matrix $\mathbf{\Sigma(\xi)}$ are
\begin{equation}
\mathbf{\rho}(t_{ij},t_{ik}) = \exp(-\mathbf{\xi}|j-k|), \quad 1\leq j<k\leq n_i.
\end{equation}
Here we consider skew-$t$ and skew-normal copula along with their symmetric counterparts for the dependence structure of the models. We set $\mathbf{\xi} = 0.25$ for all the copulas. For the skew-elliptical ones we set $\bar{\mathbf{\lambda}} = 1$ under equi-skewness, and fixed values for the degrees of freedom parameter as $\mathbf{\nu} = \{3,8,15\}$. For each simulation we generate $N = 500$ Monte Carlo samples and then estimate the parameters and their associated standard errors.
\begin{table}[]
    \centering
    \begin{small}
    \scalebox{1.0}{
    \tabcolsep = 0.18cm
    \begin{tabular}{|c|c|c c c c c c c c|}
    \hline
    & \textbf{Parameters} & $\beta_0$ & $\beta_1$ & $\beta_2$ & $\beta_3$ & $V[b]$ & $\kappa$ & $\mathbf{\xi}$ & $\bar{\mathbf{\lambda}}$ \\
    & \textbf{True Value} & 1.5 & 0.5 & 0.5 & 1.0 & 1.0 & 3.0 & 0.25 & 1.0 \\
    \hline
    \multicolumn{1}{|c}{\textbf{Copula}} & \multicolumn{1}{|c|}{Skew-t,$\mathbf{\nu} = 3$} & \multicolumn{8}{c|}{} \\
    \hline
    \textbf{m} = 200 & Mean & 1.4917 & 0.5054 & 0.4967 & 1.0001 & 0.9890 & 3.3965 & 0.3171 & 0.6465 \\
    & Bias & -0.0083 & 0.0054 & -0.0033 & 0.0001 & -0.0110 & 0.3965 & 0.0671 & -0.3535 \\
    & SD & 0.2598 & 0.1542 & 0.1525 & 0.0208 & 0.1820 & 1.1862 & 0.1192 & 0.3294 \\
    & SE & 0.2800 & 0.1850 & 0.1971 & 0.0317 & 0.2002 & 1.9486 & 0.0423 & 0.2934 \\
    & RMSE & 0.2599 & 0.1543 & 0.1525 & 0.0208 & 0.1823 & 1.2507 & 0.1368 & 0.4832 \\
    \hline
    \textbf{m} = 500 & Mean & 1.4921 & 0.5068 & 0.4953 & 0.9999 & 0.9929 & 3.2168 & 0.2848 & 0.7376 \\
    & Bias & -0.0079 & 0.0068 & -0.0047 & -0.0001 & -0.0071 & 0.2168 & 0.0348 & -0.2624 \\
    & SD & 0.1712 & 0.1003 & 0.1025 & 0.0117 & 0.1211 & 0.7187 & 0.0738 & 0.2283 \\
    & SE & 0.1680 & 0.0934 & 0.1107 & 0.0158 & 0.1064 & 0.6521 & 0.0249 & 0.2426 \\
    & RMSE & 0.1714 & 0.1005 & 0.1026 & 0.0117 & 0.1213 & 0.7507 & 0.0816 & 0.3478 \\
    \hline
    \multicolumn{1}{|c}{\textbf{Copula}} & \multicolumn{1}{|c|}{Skew-t,$\mathbf{\nu} = 8$} & \multicolumn{8}{c|}{} \\
    \hline
    \textbf{m} = 200 & Mean & 1.5115 & 0.4933 & 0.5020 & 1.0024 & 0.9678 & 3.3293 & 0.3307 & 0.6405 \\
    & Bias & 0.0115 & -0.0067 & 0.0020 & 0.0024 & -0.0322 & 0.3293 & 0.0807 & -0.3595 \\
    & SD & 0.2687 & 0.1610 & 0.1574 & 0.0188 & 0.1849 & 1.2072 & 0.1410 & 0.3908 \\
    & SE & 0.2829 & 0.1336 & 0.1754 & 0.0310 & 0.1458 & 1.5335 & 0.0416 & 0.3185 \\
    & RMSE & 0.2689 & 0.1611 & 0.1575 & 0.0189 & 0.1878 & 1.2513 & 0.1625 & 0.5310 \\
    \hline
    \textbf{m} = 500 & Mean & 1.4945 & 0.5025 & 0.4978 & 1.0008 & 1.0003 & 3.2085 & 0.3087 & 0.6910 \\
    & Bias & -0.0055 & 0.0025 & -0.0022 & 0.0008 & 0.0003 & 0.2085 & 0.0587 & -0.3090 \\
    & SD & 0.1700 & 0.1014 & 0.1003 & 0.0124 & 0.1282 & 0.7294 & 0.0886 & 0.2289 \\
    & SE & 0.1621 & 0.0773 & 0.1069 & 0.0178 & 0.0804 & 0.8080 & 0.0251 & 0.2005 \\
    & RMSE & 0.1700 & 0.1014 & 0.1004 & 0.0125 & 0.1282 & 0.7587 & 0.1063 & 0.3845 \\
    \hline
    \multicolumn{1}{|c}{\textbf{Copula}} & \multicolumn{1}{|c|}{Skew-t,$\mathbf{\nu} = 15$} & \multicolumn{8}{c|}{} \\
    \hline
    \textbf{m} = 200 & Mean & 1.5158 & 0.4904 & 0.4935 & 0.9989 & 0.9712 & 3.3751 & 0.3432 & 0.6272 \\
    & Bias & 0.0158 & -0.0096 & -0.0065 & -0.0011 & -0.0288 & 0.3751 & 0.0932 & -0.3728 \\
    & SD & 0.2655 & 0.1546 & 0.1517 & 0.0198 & 0.1884 & 1.2154 & 0.1563 & 0.4076 \\
    & SE & 0.2718 & 0.1504 & 0.1751 & 0.0310 & 0.1509 & 1.7044 & 0.0426 & 0.3271 \\
    & RMSE & 0.2659 & 0.1549 & 0.1519 & 0.0198 & 0.1906 & 1.2719 & 0.1810 & 0.5524 \\
    \hline
    \textbf{m} = 500 & Mean & 1.4984 & 0.5003 & 0.5015 & 0.9997 & 0.9994 & 3.1865 & 0.3086 & 0.6926 \\
    & Bias & -0.0016 & 0.0003 & 0.0015 & -0.0003 & -0.0006 & 0.1865 & 0.0586 & -0.3074 \\
    & SD & 0.1711 & 0.0981 & 0.0982 & 0.0128 & 0.1272 & 0.6829 & 0.0873 & 0.2690 \\
    & SE & 0.1573 & 0.0742 & 0.1000 & 0.0158 & 0.0785 & 0.6281 & 0.0248 & 0.2171 \\
    & RMSE & 0.1711 & 0.0981 & 0.0982 & 0.0128 & 0.1272 & 0.7080 & 0.1052 & 0.4085 \\
    \hline
    \multicolumn{1}{|c}{\textbf{Copula}} & \multicolumn{1}{|c|}{Skew-normal} & \multicolumn{8}{c|}{} \\
    \hline
    \textbf{m} = 200 & Mean & 1.4854 & 0.5080 & 0.5098 & 1.0024 & 0.9750 & 3.3101 & 0.3554 & 0.5956 \\
    & Bias & -0.0146 & 0.0080 & 0.0098 & 0.0024 & -0.0250 & 0.3101 & 0.1054 & -0.4044 \\
    & SD & 0.2629 & 0.1546 & 0.1563 & 0.0210 & 0.1843 & 1.1463 & 0.1621 & 0.3980 \\
    & SE & 0.2920 & 0.1406 & 0.2132 & 0.0387 & 0.1661 & 1.5973 & 0.0472 & 0.3602 \\
    & RMSE & 0.2633 & 0.1548 & 0.1565 & 0.0212 & 0.1860 & 1.1876 & 0.1933 & 0.5674 \\
    \hline
    \textbf{m} = 500 & Mean & 1.4891 & 0.5039 & 0.5017 & 1.0001 & 0.9884 & 3.1932 & 0.3211 & 0.6771 \\
    & Bias & -0.0109 & 0.0039 & 0.0017 & 0.0001 & -0.0116 & 0.1932 & 0.0711 & -0.3229 \\
    & SD & 0.1765 & 0.1009 & 0.0948 & 0.0121 & 0.1249 & 0.7273 & 0.1005 & 0.2788 \\
    & SE & 0.1576 & 0.0735 & 0.0990 & 0.0157 & 0.0791 & 0.7316 & 0.0258 & 0.2033 \\
    & RMSE & 0.1768 & 0.1009 & 0.0949 & 0.0121 & 0.1255 & 0.7525 & 0.1231 & 0.4266 \\
    \hline
    \end{tabular}}
    \end{small}
    \caption{Parameter estimation using IFM method when the marginals are distributed as Gamma. Performance for $500$ replications with skew-$t$ and skew-normal copula.}
    \label{tab:simulation1}
\end{table}
\begin{table}[]
    \centering
    \begin{small}
    \scalebox{1.0}{
    \tabcolsep = 0.18cm
    \begin{tabular}{|c|c|c c c c c c c c|}
    \hline
    & \textbf{Parameters} & $\beta_0$ & $\beta_1$ & $\beta_2$ & $\beta_3$ & $V[b]$ & $\kappa$ & $\mathbf{\xi}$ & $\bar{\mathbf{\lambda}}$ \\
    & \textbf{True Value} & 1.5 & 0.5 & 0.5 & 1.0 & 1.0 & 3.0 & 0.25 & - \\
    \hline
    \multicolumn{1}{|c}{\textbf{Copula}} & \multicolumn{1}{|c|}{Student-t,$\mathbf{\nu} = 3$} & \multicolumn{8}{c|}{} \\
    \hline
    \textbf{m} = 200 & Mean & 1.4864 & 0.5049 & 0.5004 & 1.0002 & 0.9782 & 3.5281 & 0.2811 & - \\
    & Bias & -0.0136 & 0.0049 & 0.0004 & 0.0002 & -0.0218 & 0.5281 & 0.0311 & - \\
    & SD & 0.2704 & 0.1567 & 0.1644 & 0.0169 & 0.1831 & 1.4162 & 0.1049 & - \\
    & SE & 0.2963 & 0.2271 & 0.2142 & 0.0441 & 0.2264 & 2.1010 & 0.0243 & - \\
    & RMSE & 0.2708 & 0.1568 & 0.1644 & 0.0169 & 0.1844 & 1.5114 & 0.1094 & - \\
    \hline
    \textbf{m} = 500 & Mean & 1.4922 & 0.4982 & 0.5029 & 0.9999 & 1.0026 & 3.3119 & 0.2702 & - \\
    & Bias & -0.0078 & -0.0018 & 0.0029 & -0.0001 & 0.0026 & 0.3119 & 0.0202 & - \\
    & SD & 0.1725 & 0.0997 & 0.1014 & 0.0098 & 0.1354 & 0.8041 & 0.0646 & - \\
    & SE & 0.1945 & 0.0791 & 0.1323 & 0.0144 & 0.0895 & 1.3090 & 0.0153 & - \\
    & RMSE & 0.1727 & 0.0997 & 0.1015 & 0.0098 & 0.1355 & 0.8625 & 0.0677 & - \\
    \hline
    \multicolumn{1}{|c}{\textbf{Copula}} & \multicolumn{1}{|c|}{Student-t,$\mathbf{\nu} = 8$} & \multicolumn{8}{c|}{} \\
    \hline
    \textbf{m} = 200 & Mean & 1.5173 & 0.4930 & 0.4892 & 0.9996 & 0.9842 & 3.4324 & 0.2822 & - \\
    & Bias & 0.0173 & -0.0070 & -0.0108 & -0.0004 & -0.0158 & 0.4324 & 0.0322 & - \\
    & SD & 0.2819 & 0.1692 & 0.1582 & 0.0160 & 0.1904 & 1.2955 & 0.1074 & - \\
    & SE & 0.3126 & 0.1634 & 0.2097 & 0.0320 & 0.1812 & 2.1569 & 0.0223 & - \\
    & RMSE & 0.2824 & 0.1694 & 0.1585 & 0.0160 & 0.1911 & 1.3658 & 0.1122 & - \\
    \hline
    \textbf{m} = 500 & Mean & 1.5118 & 0.4953 & 0.4958 & 0.9998 & 0.9939 & 3.1866 & 0.2636 & - \\
    & Bias & 0.0118 & -0.0047 & -0.0042 & -0.0002 & -0.0061 & 0.1866 & 0.0137 & - \\
    & SD & 0.1730 & 0.0979 & 0.1017 & 0.0105 & 0.1411 & 0.7289 & 0.0670 & - \\
    & SE & 0.1649 & 0.0773 & 0.1035 & 0.0138 & 0.1118 & 0.6927 & 0.0135 & - \\
    & RMSE & 0.1734 & 0.0980 & 0.1017 & 0.0106 & 0.1413 & 0.7524 & 0.0683 & - \\
    \hline
    \multicolumn{1}{|c}{\textbf{Copula}} & \multicolumn{1}{|c|}{Student-t,$\mathbf{\nu} = 15$} & \multicolumn{8}{c|}{} \\
    \hline
    \textbf{m} = 200 & Mean & 1.4785 & 0.5077 & 0.5058 & 1.0004 & 0.9943 & 3.5381 & 0.2953 & - \\
    & Bias & -0.0215 & 0.0077 & 0.0058 & 0.0004 & -0.0057 & 0.5381 & 0.0453 & - \\
    & SD & 0.2665 & 0.1613 & 0.1647 & 0.0171 & 0.1843 & 1.2716 & 0.1122 & - \\
    & SE & 0.2860 & 0.1512 & 0.1967 & 0.0343 & 0.1752 & 2.2277 & 0.0223 & - \\
    & RMSE & 0.2673 & 0.1615 & 0.1648 & 0.0171 & 0.1844 & 1.3808 & 0.1210 & - \\
    \hline
    \textbf{m} = 500 & Mean & 1.4941 & 0.5032 & 0.4931 & 0.9999 & 0.9930 & 3.2570 & 0.2716 & - \\
    & Bias & -0.0059 & 0.0032 & -0.0069 & -0.0001 & -0.0070 & 0.2570 & 0.0216 & - \\
    & SD & 0.1788 & 0.1029 & 0.1065 & 0.0102 & 0.1330 & 0.7906 & 0.0715 & - \\
    & SE & 0.1683 & 0.0818 & 0.1083 & 0.0142 & 0.0837 & 0.7635 & 0.0132 & - \\
    & RMSE & 0.1789 & 0.1029 & 0.1067 & 0.0102 & 0.1332 & 0.8313 & 0.0748 & - \\
    \hline
    \multicolumn{1}{|c}{\textbf{Copula}} & \multicolumn{1}{|c|}{Gaussian} & \multicolumn{8}{c|}{} \\
    \hline
    \textbf{m} = 200 & Mean & 1.4976 & 0.4988 & 0.4987 & 0.9999 & 0.9722 & 3.4854 & 0.2925 & - \\
    & Bias & -0.0024 & -0.0012 & -0.0013 & -0.0001 & -0.0278 & 0.4854 & 0.0425 & - \\
    & SD & 0.2643 & 0.1597 & 0.1660 & 0.0163 & 0.1953 & 1.3667 & 0.1264 & - \\
    & SE & 0.2879 & 0.1429 & 0.2174 & 0.0337 & 0.1844 & 1.7209 & 0.0209 & - \\
    & RMSE & 0.2643 & 0.1597 & 0.1660 & 0.0163 & 0.1973 & 1.4504 & 0.1333 & - \\
    \hline
    \textbf{m} = 500 & Mean & 1.4985 & 0.5006 & 0.4991 & 1.0000 & 0.4937 & 3.2284 & 0.2725 & - \\
    & Bias & -0.0015 & 0.0006 & -0.0009 & 0.0000 & -0.0063 & 0.2284 & 0.0225 & - \\
    & SD & 0.1780 & 0.1039 & 0.1025 & 0.0096 & 0.1320 & 0.7880 & 0.0778 & - \\
    & SE & 0.1616 & 0.0766 & 0.1030 & 0.0142 & 0.0812 & 0.6222 & 0.0125 & - \\
    & RMSE & 0.1781 & 0.1040 & 0.1025 & 0.0096 & 0.1321 & 0.8204 & 0.0809 & - \\
    \hline
    \end{tabular}}
    \end{small}
    \caption{Parameter estimation using IFM method when the marginals are distributed as Gamma. Performance for $500$ replications with Student-$t$ and Gaussian copula.}
    \label{tab:simulation2}
\end{table}
\begin{table}[]
    \centering
    \begin{small}
    \scalebox{1.0}{
    \tabcolsep = 0.18cm
    \begin{tabular}{|c|c|c c c c c c c c|}
    \hline
    & \textbf{Parameters} & $\beta_0$ & $\beta_1$ & $\beta_2$ & $\beta_3$ & $V[b]$ & $\sigma$ & $\mathbf{\xi}$ & $\bar{\mathbf{\lambda}}$ \\
    & \textbf{True Value} & 1.5 & 0.5 & 0.5 & 1.0 & 1.0 & 1.0 & 0.25 & 1.0 \\
    \hline
    \multicolumn{1}{|c}{\textbf{Copula}} & \multicolumn{1}{|c|}{Skew-t,$\mathbf{\nu} = 3$} & \multicolumn{8}{c|}{} \\
    \hline
    \textbf{m} = 200 & Mean & 1.4713 & 0.5115 & 0.5102 & 1.0002 & 1.0388 & 0.9768 & 0.2753 & 0.6528 \\
    & Bias & -0.0287 & 0.0115 & 0.0102 & 0.0002 & 0.0388 & -0.0232 & 0.0253 & -0.3472 \\
    & SD & 0.3010 & 0.1882 & 0.1765 & 0.0307 & 0.1674 & 0.1411 & 0.0965 & 0.2841 \\
    & SE & 0.2722 & 0.1300 & 0.1744 & 0.0308 & 0.2206 & 0.1874 & 0.0694 & 0.2932 \\
    & RMSE & 0.3113 & 0.1886 & 0.1768 & 0.0307 & 0.1736 & 0.1430 & 0.1000 & 0.4486 \\
    \hline
    \textbf{m} = 500 & Mean & 1.4886 & 0.5039 & 0.5071 & 0.9990 & 1.0370 & 0.9771 & 0.2744 & 0.6586 \\
    & Bias & -0.0114 & 0.0039 & 0.0071 & -0.0010 & 0.0374 & -0.0229 & 0.0244 & -0.3414 \\
    & SD & 0.1929 & 0.1174 & 0.1139 & 0.0210 & 0.1055 & 0.0756 & 0.0656 & 0.1700 \\
    & SE & 0.1694 & 0.0818 & 0.1106 & 0.0196 & 0.0947 & 0.0677 & 0.0447 & 0.1784 \\
    & RMSE & 0.1933 & 0.1174 & 0.1141 & 0.0210 & 0.1120 & 0.0720 & 0.0700 & 0.3901 \\
    \hline
    \multicolumn{1}{|c}{\textbf{Copula}} & \multicolumn{1}{|c|}{Skew-t,$\mathbf{\nu} = 8$} & \multicolumn{8}{c|}{} \\
    \hline
    \textbf{m} = 200 & Mean & 1.4944 & 0.5033 & 0.5101 & 0.9990 & 1.0370 & 0.9771 & 0.2771 & 0.6282 \\
    & Bias & -0.0056 & 0.0033 & 0.0101 & -0.0010 & 0.0370 & -0.0229 & 0.0271 & -0.3718 \\
    & SD & 0.2751 & 0.1728 & 0.1711 & 0.0311 & 0.1583 & 0.1320 & 0.0920 & 0.3209 \\
    & SE & 0.2697 & 0.1278 & 0.1714 & 0.0307 & 0.2076 & 0.1965 & 0.0720 & 0.3146 \\
    & RMSE & 0.2752 & 0.1729 & 0.1714 & 0.0311 & 0.1626 & 0.1340 & 0.0959 & 0.4911 \\
    \hline
    \textbf{m} = 500 & Mean & 1.5046 & 0.4976 & 0.5020 & 1.0002 & 1.0349 & 0.9784 & 0.2764 & 0.6915 \\
    & Bias & 0.0046 & -0.0024 & 0.0020 & 0.0002 & 0.0349 & -0.0216 & 0.0264 & -0.3085 \\
    & SD & 0.1859 & 0.1120 & 0.1107 & 0.0189 & 0.1059 & 0.0989 & 0.0520 & 0.1556 \\
    & SE & 0.1689 & 0.0812 & 0.1094 & 0.0195 & 0.0943 & 0.0762 & 0.0382 & 0.1449 \\ 
    & RMSE & 0.1860 & 0.1120 & 0.1107 & 0.0189 & 0.1115 & 0.1012 & 0.0583 & 0.3455 \\
    \hline
    \multicolumn{1}{|c}{\textbf{Copula}} & \multicolumn{1}{|c|}{Skew-t,$\mathbf{\nu} = 15$} & \multicolumn{8}{c|}{} \\
    \hline
    \textbf{m} = 200 & Mean & 1.5035 & 0.4960 & 0.5011 & 1.0018 & 1.0370 & 0.9773 & 0.2767 & 0.6409 \\
    & Bias & 0.0035 & -0.0040 & 0.0011 & 0.0018 & 0.0370 & -0.0227 & 0.0267 & -0.3591 \\
    & SD & 0.2921 & 0.1747 & 0.1734 & 0.0292 & 0.1594 & 0.1292 & 0.0801 & 0.2943 \\
    & SE & 0.2663 & 0.1274 & 0.1721 & 0.0311 & 0.2256 & 0.1952 & 0.0670 & 0.2516 \\
    & RMSE & 0.2922 & 0.1747 & 0.1734 & 0.0293 & 0.1636 & 0.1312 & 0.0844 & 0.4643 \\
    \hline
    \textbf{m} = 500 & Mean & 1.4981 & 0.5037 & 0.5009 & 1.0018 & 1.0367 & 0.9775 & 0.2767 & 0.6887 \\
    & Bias & -0.0019 & 0.0037 & 0.0009 & 0.0010 & 0.0367 & -0.0225 & 0.0267 & -0.3113 \\
    & SD & 0.1835 & 0.1102 & 0.1057 & 0.0200 & 0.1042 & 0.0887 & 0.0513 & 0.1672 \\
    & SE & 0.1695 & 0.0815 & 0.1097 & 0.0195 & 0.0927 & 0.0723 & 0.0363 & 0.1457 \\
    & RMSE & 0.1845 & 0.1105 & 0.1057 & 0.0200 & 0.1105 & 0.0915 & 0.0578 & 0.3534 \\
    \hline
    \multicolumn{1}{|c}{\textbf{Copula}} & \multicolumn{1}{|c|}{Skew-normal} & \multicolumn{8}{c|}{} \\
    \hline
    \textbf{m} = 200 & Mean & 1.4982 & 0.4987 & 0.5097 & 0.9986 & 1.0386 & 0.9775 & 0.2768 & 0.6307 \\
    & Bias & -0.0018 & -0.0013 & 0.0097 & -0.0014 & 0.0386 & -0.0225 & 0.0268 & -0.3693 \\
    & SD & 0.2889 & 0.1750 & 0.1821 & 0.0319 & 0.1587 & 0.1275 & 0.0791 & 0.3435 \\
    & SE & 0.2704 & 0.1292 & 0.1723 & 0.0309 & 0.2309 & 0.2162 & 0.0650 & 0.2810 \\
    & RMSE & 0.2889 & 0.1750 & 0.1824 & 0.0319 & 0.1633 & 0.1295 & 0.0835 & 0.5044 \\
    \hline
    \textbf{m} = 500 & Mean & 1.4984 & 0.4990 & 0.5021 & 0.9997 & 1.0381 & 0.9977 & 0.2764 & 0.6963  \\
    & Bias & -0.0016 & -0.0010 & 0.0021 & -0.0003 & 0.0381 & -0.0223 & 0.0264 & -0.3037 \\
    & SD & 0.1911 & 0.1156 & 0.1109 & 0.0196 & 0.0981 & 0.0859 & 0.0451 & 0.1563 \\
    & SE & 0.1708 & 0.0839 & 0.1099 & 0.0195 & 0.0818 & 0.0745 & 0.0332 & 0.1509 \\
    & RMSE & 0.1911 & 0.1156 & 0.1109 & 0.0196 & 0.1052 & 0.0887 & 0.0523 & 0.3416 \\
    \hline
    \end{tabular}}
    \end{small}
    \caption{Parameter estimation using IFM method when the marginals are distributed as normal. Performance for $500$ replications with skew-$t$ and skew-normal copula.}
    \label{tab:simulation3}
\end{table}
\begin{table}[]
    \centering
    \begin{small}
    \scalebox{1.0}{
    \tabcolsep = 0.18cm
    \begin{tabular}{|c|c|c c c c c c c c|}
    \hline
    & \textbf{Parameters} & $\beta_0$ & $\beta_1$ & $\beta_2$ & $\beta_3$ & $V[b]$ & $\sigma$ & $\mathbf{\xi}$ & $\bar{\mathbf{\lambda}}$ \\
    & \textbf{True Value} & 1.5 & 0.5 & 0.5 & 1.0 & 1.0 & 1.0 & 0.25 & - \\
    \hline
    \multicolumn{1}{|c}{\textbf{Copula}} & \multicolumn{1}{|c|}{Student-t,$\mathbf{\nu} = 3$} & \multicolumn{8}{c|}{} \\
    \hline
    \textbf{m} = 200 & Mean & 1.4715 & 0.5208 & 0.4996 & 1.0014 & 1.0560 & 0.9627 & 0.2855 & - \\
    & Bias & -0.0285 & 0.0208 & -0.0004 & 0.0014 & 0.0560 & -0.0373 & 0.0355 & - \\
    & SD & 0.3126 & 0.1852 & 0.1884 & 0.0258 & 0.1862 & 0.1124 & 0.1471 & - \\
    & SE & 0.3352 & 0.2044 & 0.2088 & 0.0621 & 0.2129 & 0.0998 & 0.0887 & - \\
    & RMSE & 0.3139 & 0.1864 & 0.1884 & 0.0258 & 0.1944 & 0.1184 & 0.1513 & - \\
    \hline
    \textbf{m} = 500 & Mean & 1.5017 & 0.4986 & 0.4995 & 0.9989 & 1.0584 & 0.9635 & 0.2837 & - \\
    & Bias & 0.0017 & -0.0014 & -0.0005 & -0.0011 & 0.0584 & -0.0365 & 0.0337 & - \\
    & SD & 0.2007 & 0.1200 & 0.1176 & 0.0158 & 0.1242 & 0.0804 & 0.1020 & - \\
    & SE & 0.1982 & 0.0968 & 0.1307 & 0.0229 & 0.1476 & 0.0507 & 0.0487 & - \\
    & RMSE & 0.2007 & 0.1200 & 0.1176 & 0.0158 & 0.1372 & 0.0883 & 0.1074 & - \\
    \hline
    \multicolumn{1}{|c}{\textbf{Copula}} & \multicolumn{1}{|c|}{Student-t,$\mathbf{\nu} = 8$} & \multicolumn{8}{c|}{} \\
    \hline
    \textbf{m} = 200 & Mean & 1.4872 & 0.5041 & 0.5054 & 0.9991 & 1.0557 & 0.9630 & 0.2909 & - \\
    & Bias & -0.0128 & 0.0041 & 0.0054 & -0.0009 & 0.0557 & -0.0370 & 0.0409 & - \\
    & SD & 0.3053 & 0.1881 & 0.1871 & 0.0256 & 0.1716 & 0.1125 & 0.1427 & - \\
    & SE & 0.3361 & 0.2035 & 0.2067 & 0.0608 & 0.2111 & 0.1333 & 0.0888 & - \\ 
    & RMSE & 0.3056 & 0.1881 & 0.1872 & 0.0256 & 0.1804 & 0.1184 & 0.1484 & - \\
    \hline
    \textbf{m} = 500 & Mean & 1.5067 & 0.4963 & 0.5041 & 0.9994 & 1.0495 & 0.9633 & 0.2861 & - \\
    & Bias & 0.0115 & -0.0027 & 0.0041 & -0.0006 & 0.0495 & -0.0361 & 0.0361 & - \\
    & SD & 0.1918 & 0.1211 & 0.1211 & 0.0154 & 0.1214 & 0.0816 & 0.0978 & - \\
    & SE & 0.1985 & 0.0951 & 0.1273 & 0.0241 & 0.1126 & 0.0549 & 0.0427 & - \\
    & RMSE & 0.1923 & 0.1212 & 0.1212 & 0.0156 & 0.1311 & 0.0892 & 0.1042 & - \\
    \hline
    \multicolumn{1}{|c}{\textbf{Copula}} & \multicolumn{1}{|c|}{Student-t,$\mathbf{\nu} = 15$} & \multicolumn{8}{c|}{} \\
    \hline
    \textbf{m} = 200 & Mean & 1.5067 & 0.5052 & 0.4917 & 0.9998 & 1.0026 & 0.9814 & 0.2643 & - \\
    & Bias & 0.0067 & 0.0052 & -0.0083 & -0.0002 & 0.0026 & -0.0186 & 0.0143 & - \\
    & SD & 0.3109 & 0.1900 & 0.1893 & 0.0238 & 0.1568 & 0.1200 & 0.1414 & - \\
    & SE & 0.3356 & 0.1810 & 0.1921 & 0.0256 & 0.1869 & 0.1312 & 0.0799 & - \\
    & RMSE & 0.3109 & 0.1901 & 0.1895 & 0.0238 & 0.1570 & 0.1214 & 0.1429 & - \\
    \hline
    \textbf{m} = 500 & Mean & 1.4939 & 0.5044 & 0.5021 & 0.9998 & 1.0022 & 0.9908 & 0.2567 & - \\
    & Bias & -0.0061 & 0.0044 & 0.0021 & -0.0002 & 0.0022 & -0.0092 & 0.0067 & - \\
    & SD & 0.1937 & 0.1203 & 0.1146 & 0.0153 & 0.1076 & 0.0784 & 0.0955 & - \\
    & SE & 0.1748 & 0.0843 & 0.1134 & 0.0154 & 0.0889 & 0.0561 & 0.0433 & - \\
    & RMSE & 0.1938 & 0.1203 & 0.1147 & 0.0153 & 0.1078 & 0.0786 & 0.0957 & - \\
    \hline
    \multicolumn{1}{|c}{\textbf{Copula}} & \multicolumn{1}{|c|}{Gaussian} & \multicolumn{8}{c|}{} \\
    \hline
    \textbf{m} = 200 & Mean & 1.4982 & 0.5082 & 0.4927 & 0.9993 & 1.0066 & 0.9836 & 0.2633 & - \\
    & Bias & -0.0018 & 0.0082 & -0.0073 & -0.0007 & 0.0066 & -0.0164 & 0.0133 & - \\
    & SD & 0.3053 & 0.1885 & 0.1860 & 0.0233 & 0.1512 & 0.1102 & 0.1377 & - \\
    & SE & 0.3153 & 0.1315 & 0.1775 & 0.0293 & 0.1732 & 0.1242 & 0.0751 & - \\
    & RMSE & 0.3053 & 0.1885 & 0.1861 & 0.0233 & 0.1514 & 0.1114 & 0.1383 & - \\
    \hline
    \textbf{m} = 500 & Mean & 1.4988 & 0.5070 & 0.5038 & 1.0010 & 1.0060 & 0.9928 & 0.2557 & - \\
    & Bias & -0.0012 & 0.0070 & 0.0038 & 0.0010 & 0.0060 & -0.0072 & 0.0057 & - \\
    & SD & 0.2047 & 0.1255 & 0.1210 & 0.0160 & 0.1037 & 0.0777 & 0.0951 & - \\
    & SE & 0.1754 & 0.0842 & 0.1135 & 0.0153 & 0.0843 & 0.0543 & 0.0447 & - \\
    & RMSE & 0.2051 & 0.1256 & 0.1210 & 0.0160 & 0.1039 & 0.0780 & 0.0953 & - \\
    \hline
    \end{tabular}}
    \end{small}
    \caption{Parameter estimation using IFM method when the marginals are distributed as normal. Performance for $500$ replications with Student-$t$ and Gaussian copula.}
    \label{tab:simulation4}
\end{table}
\par Table \ref{tab:simulation1} and \ref{tab:simulation2} show parameter estimations of the class of models in (\ref{simmodel1}) using skew-elliptical and elliptical copulas and gamma marginals with Gaussian random effects. We present the mean, the biases [$\frac{1}{N} \sum_{i=1}^N (\hat{\theta}^*_j - \theta^*)$], roots of mean square errors [$\sqrt{\frac{1}{N} \sum_{i=1}^N (\hat{\theta}^*_j - \theta^*)^2}$], empirical standard errors (denoted as SD) and the standard errors obtained from the asymptotic covariance matrices (denoted as SE), where $\hat{\theta}^*_j$ is the parameter estimates for the $j$-th sample. In Table \ref{tab:simulation3} and \ref{tab:simulation4}, we provide additional simulation results using normal marginals as well. The results show that parameter estimations tend to have higher accuracies based on larger sample size $m$ with smaller Bias, SE and RMSE. However, there is systematic bias in the estimation of the skewness parameter $\bar{\mathbf{\lambda}}$ for all the models, and the shape parameter $\kappa$ for the Gamma based models. We notice that using IFM method does not change the precision of estimation of the marginal parameters very much. Overall, the estimation of parameters are accurate and SE and SD are relatively close to each other, which shows this estimating approach is viable for drawing inference from real data set.
\section{Data analysis}
Due to the skewed nature of the CD4 count marker with positive real support, we aim to model the marginals using a Gamma mixed model and the temporal dependence using skewed multivariate copulas to assess disease progression. Our model explores changes in CD4 counts over time within patients. Despite attempts to transform the data using logarithmic or square root transformations, the skewness persists. To facilitate estimation and interpretation of coefficients, we apply a scale transformation to the CD4$^+$ T cell counts by dividing them by $100$. We observed sparse data in the $5$-th visit column and hence omit it from our analysis. Additionally, some entries were missing in the fourth visit column, and we imputed these using the carry-forward method as discussed in \citet{suresh2021gaussian}. Our predictors include age, gender, first baseline regimen, and initial weight of each patient. Gender is represented using indicators: 0 for female and 1 for male patients. The first baseline regimen (FBR) of ARV combination is encoded as $1, 2$, or $5$. Therefore, referring to the models in (\ref{model1}) we consider -
\begin{equation}\label{realmodel}
x_{ij}\beta + d_{ij}\mathbf{b}_i = \beta_0 + \mathrm{gender}_i\beta_1 + \mathrm{age}_i\beta_2 + \mathrm{fbr}_i\beta_3 + \mathrm{weight}_i\beta_4 + t_{ij}\beta_5 + b_i,
\end{equation}
and $Y_{ij}$ is the CD4 count at $j$-th time point for the $i$-th patient (normalized by $100$). The time variable is rescaled as $t_{ij} = (\mathrm{week} - 18)/12$. We have considered random intercept structure in the models. Based on the sample correlation matrix of this data set, $AR(1)$ structure for the correlation matrices for the multivariate copulas seems to be appropriate. After the rescaling of the time points, the entries of the correlation matrix $\mathbf{\Sigma_i}$, are equivalent to $\mathbf{\rho}(t_{ij},t_{ik}) = \exp(-\mathbf{\xi}|j - k|),\;\; 1\leq k < j \leq n_i$. Considering two marginal mixed models with four multivariate copulas, we estimate the parameters using the method described in Section \ref{sec5}. Since we use fixed integer valued $\mathbf{\nu}$ in the skew-$t$ and Student-$t$ copula, we select the value with in the set $\{3,\dots,30\}$ based on the maximum value of the log-likelihood.
\begin{table}[]
    \centering
    \begin{small}
    \scalebox{1.0}{
    \tabcolsep = 0.18cm
    \begin{tabular}{|c|c c|c c|}
    \hline
    \multicolumn{1}{|c}{} & \multicolumn{2}{|c|}{Gamma marginals} & \multicolumn{2}{c|}{Normal marginals} \\
    \hline
    \textbf{Parameters} & Est. & SE & Est. & SE \\
    \hline
    $\beta_0$ & 0.2533 & 0.1558 & 1.3204 & 0.4558 \\
    $\beta_1$ & 0.0959 & 0.0539 & 0.1264 & 0.1454 \\
    $\beta_2$ & 0.0025 & 0.0019 & 0.0011 & 0.0049 \\
    $\beta_3$ & 0.0114 & 0.0154 & 0.0201 & 0.0408 \\
    $\beta_4$ & 0.0113 & 0.0015 & 0.0273 & 0.0042 \\
    $\beta_5$ & 0.0907 & 0.0103 & 0.2022 & 0.0269 \\
    $V[b]$ & 0.0700 & 0.0258 & 1.2140 & 0.3390 \\
    $\kappa$ & 5.0979 & 1.9562 & - & - \\
    $\sigma$ & - & - & 0.8890 & 0.1394 \\
    \hline
    \end{tabular}}
    \end{small}
    \caption{Marginal parameter estimation of HIV CD4$^+$ T cell count data with model (\ref{realmodel}) using Gamma and normal mixed model.}
    \label{tab:tablemarest1}
\end{table}
\begin{table}[]
    \centering
    \begin{small}
    \scalebox{1.0}{
    \tabcolsep = 0.18cm
    \begin{tabular}{|c|c|c c c c|}
    \hline
    \multicolumn{6}{|c|}{Gamma marginals} \\
    \hline
    \textbf{Copula} & degrees of freedom ($\mathbf{\nu}$) & 3 & 4 & 5 & 6 \\
    \hline
    Skew-$t$ & Log-likelihood & \textbf{-1250.84} & -1261.45 & -1272.17 & -1281.84 \\
    Student-$t$ & Log-likelihood & -1288.30 & -1304.63 & -1338.77 & -1347.63 \\
    \hline
    \multicolumn{6}{|c|}{Normal marginals} \\
    \hline
    \textbf{Copula} & degrees of freedom ($\mathbf{\nu}$) & 3 & 4 & 5 & 6 \\
    \hline
    Skew-$t$ & Log-likelihood & \textbf{-1256.88} & -1271.31 & -1285.62 & -1296.17 \\
    Student-$t$ & Log-likelihood & -1257.02 & -1271.50 & -1285.89 & -1297.65 \\
    \hline
    \end{tabular}}
    \end{small}
    \caption{Estimation of the degrees of freedom parameter for the skew-$t$ and Student-$t$ copula based on the maximum log-likelihood.}
    \label{tab:tabledfest1}
\end{table}
\begin{table}[]
    \centering
    \begin{small}
    \scalebox{1.0}{
    \tabcolsep=0.18cm
    \begin{tabular}{|c|c|c|c c|c|c|c|}
    \hline
    \textbf{Model} & \textbf{Copula} & \textbf{Parameters} & Est. & SE & Log-likelihood & AIC & BIC \\
    \hline
    Gamma & Skew-$t$,$\mathbf{\nu} = 3$ & $\mathbf{\xi}$ & 0.1781 & 0.0190 & \textbf{-1250.84} & \textbf{2521.67} & \textbf{2557.32} \\
    & & $\bar{\mathbf{\lambda}}$ & 1.2765 & 0.5373 & & & \\
    & Skew-normal & $\mathbf{\xi}$ & 0.1904 & 0.0329 & -1422.96 & 2863.92 & 2895.99 \\
    & & $\bar{\mathbf{\lambda}}$ & 1.8547 & 0.4033 & & & \\
    & Student-$t$,$\mathbf{\nu} = 3$ & $\mathbf{\xi}$ & 0.2052 & 0.0256 & -1288.30 & 2594.60 & 2626.68 \\
    & Gaussian & $\mathbf{\xi}$ & 0.4525 & 0.0810 & -1468.57 & 2953.14 & 2981.65 \\
    \hline
    Normal & Skew-$t$,$\mathbf{\nu} = 3$ & $\mathbf{\xi}$ & 0.2611 & 0.0285 & \textbf{-1256.88} & \textbf{2535.77} & \textbf{2594.98} \\
    & & $\bar{\mathbf{\lambda}}$ & -0.0156 & 0.0650 & & & \\
    & Skew-normal & $\mathbf{\xi}$ & 0.3084 & 0.0481 & -1429.61 & 2879.21 & 2914.86 \\
    & & $\bar{\mathbf{\lambda}}$ & -0.5016 & 0.0850 & & & \\
    & Student-$t$,$\mathbf{\nu} = 3$ & $\mathbf{\xi}$ & 0.2612 & 0.0285 & -1257.02 & 2534.04 & 2569.68 \\
    & Gaussian & $\mathbf{\xi}$ & 0.5358 & 0.1113 & -1480.53 & 2979.05 & 3011.13 \\
    \hline
    \end{tabular}}
    \end{small}
    \caption{Dependence parameter estimation of HIV CD4$^+$ T cell count data with model (\ref{realmodel}). Maximum log-likelihood value, AIC and BIC for the skew-$t$, skew-normal, Student-$t$ and Gaussian copula respectively.}
    \label{tab:tablecopest1}
\end{table}
\begin{figure}[h]
    \centering
    \includegraphics[width = 7cm]{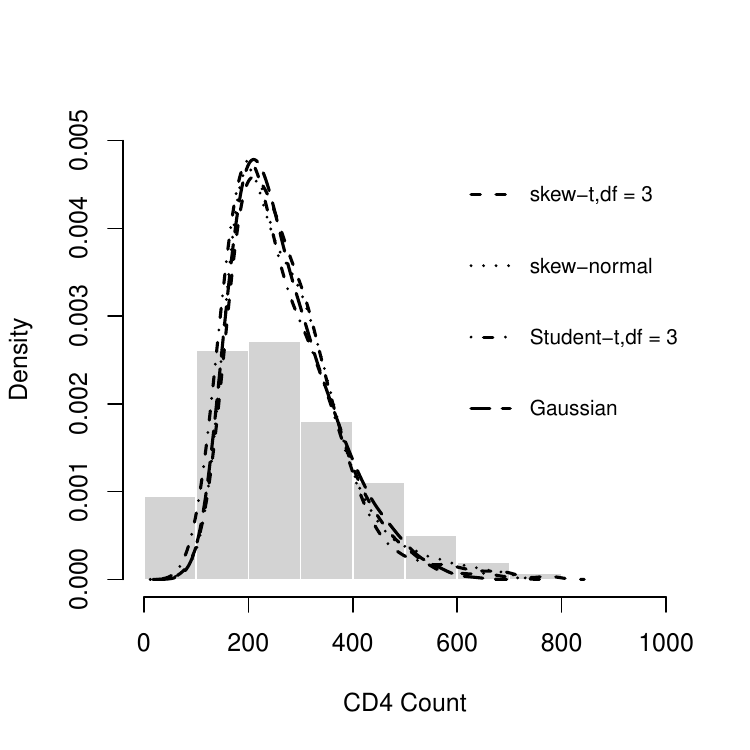}
    \includegraphics[width = 7cm]{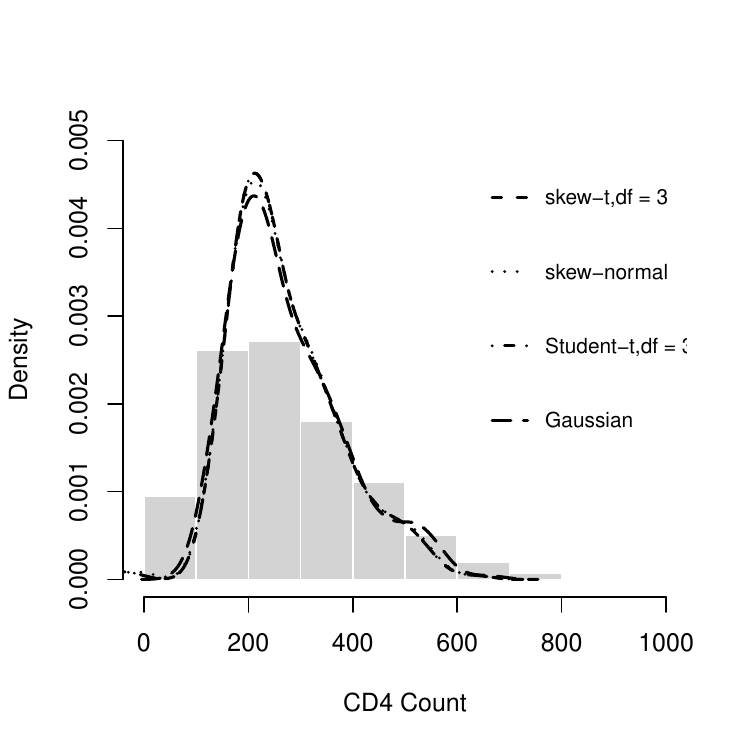}
    \caption{Fitting of HIV CD4$^+$ T cell count data with model (\ref{realmodel}) using Gamma marginals (left panel) and normal marginals (right panel). The histograms show the frequency distribution of observed CD4 counts with different dotted lines representing the fitted models.}
    \label{fig:marginalfit1}
\end{figure}
\begin{figure}[h]
    \centering
    \includegraphics[width = 16cm]{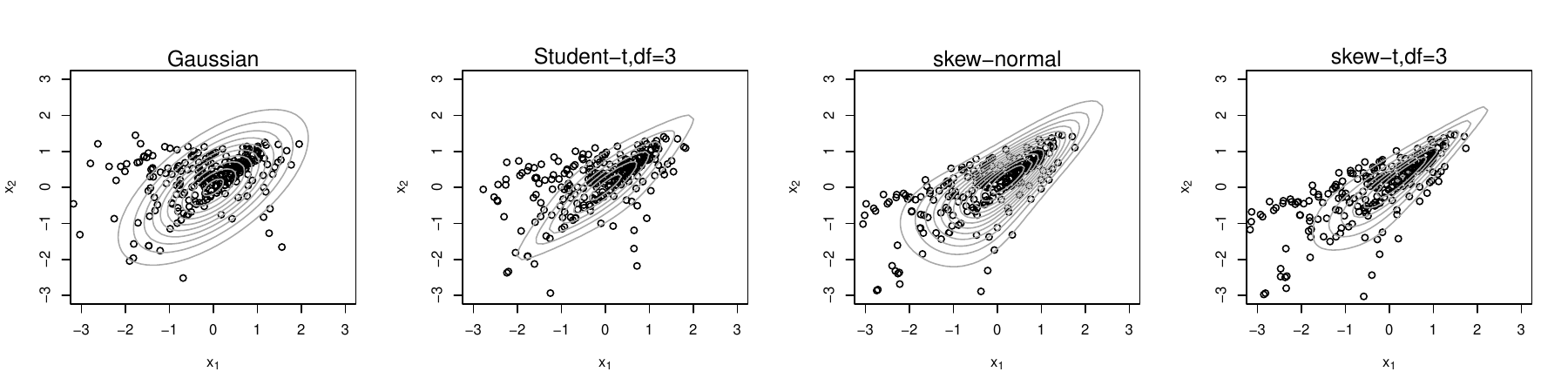}
    \includegraphics[width = 16cm]{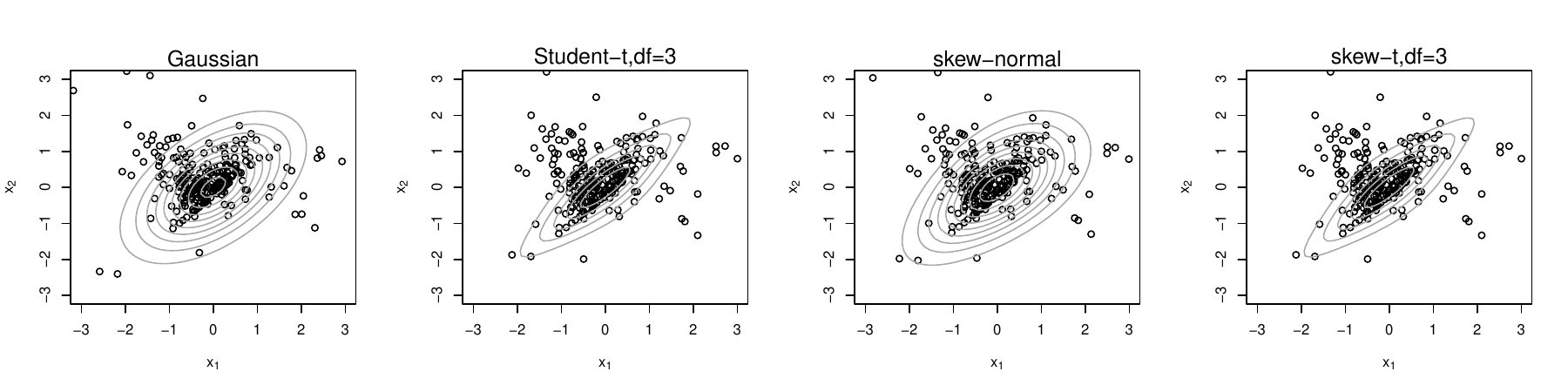}
    \caption{Fitting of the copula data (transformed to standard normal margins) using Gamma (upper panel) and normal mixed model (lower panel) of the first two time points, including the contour lines of the fitted skew and elliptical copulas, respectively.}
    \label{fig:copulafit1}
\end{figure}
\par Since the IFM estimation method yields identical marginal parameter estimates for all multivariate copula-based models, we summarize the marginal parameter estimates and their corresponding standard errors for the mixed model (\ref{realmodel}) with Gamma and normal marginals when the copula is Gaussian in Table \ref{tab:tablemarest1}. Additionally, Table \ref{tab:tabledfest1} showcases the estimation of the fixed degrees of freedom parameter for the skew-$t$ and Student-$t$ copulas based on maximum log-likelihood. Furthermore, Table \ref{tab:tablecopest1} presents the estimates of the dependence parameters along with their corresponding standard errors, log-likelihood (of the full model), AIC, and BIC for the skew-$t$, skew-normal, Student-$t$, and Gaussian copulas, respectively. We estimate the random effects from the posterior modes for each subject under different multivariate models. This allows us to visualize the models graphically when the unobserved random effects are estimated. Figure \ref{fig:marginalfit1} displays histograms of the CD4 counts of 261 HIV patients, with dotted lines representing fitted models using two marginals and four multivariate copulas, respectively. Similarly, using the estimated random effects, we transform the data into normal scores using the cumulative distribution functions of the marginals ($F(\cdot)$). In Figure \ref{fig:copulafit1}, we provide contour plots of the fitted skew and elliptical copulas to the data of the first two time points using two different marginal distributions.
\par From our analysis, it's evident that the estimates of $\beta_2$ are close to zero under both marginal models, indicating that age has a minor impact on disease progression. Conversely, the estimate of $\beta_1$ suggests that gender has a significant effect on HIV progression, as previously indicated in the profile plots in Figure \ref{fig:profiteplot1}. Additionally, the effect of time as a covariate is significant, as demonstrated by the estimate of $\beta_5$. Based on the selection criteria, we find that the skew-$t$ copula-based Gamma mixed model provides the best fit to the data among the eight candidate models. A smaller value of the degrees of freedom parameter implies stronger tail dependence across time in the data. The estimates of the skewness parameter indicate reflection and permutation asymmetry when the marginals are assumed to be Gamma. However, with normal marginals, these estimates are lower and tend towards negative values. This discrepancy may be due to the fact that modeling marginals with symmetric distributions undermines the asymmetry present in both the marginals and the dependence structure. Therefore, it's crucial to carefully choose appropriate marginals for copula-based modeling, ensuring computational tractability. Overall, the considered class of copula-based mixed models performs reasonably well, as evidenced by our ability to accurately estimate $13$ coefficients for $261$ observations.
\section{Conclusion and summary}\label{sec8}
In this paper, we explored an extension of the classical linear mixed model framework, which relies on the assumption of multivariate normality. Motivated by real-world HIV data, we proposed skew-elliptical copula-based generalized linear mixed models for analyzing such data. Our approach treats the standard linear mixed model as a special case, offering greater flexibility when the data violate the normality assumption. Our proposed class of models offers improved fit in scenarios where normality assumptions are not met, effectively capturing reflection, permutation asymmetry, and tail dependence, if present in the data. We applied our methods to model disease progression using CD4 T$^+$ cell counts from the HIV dataset. The general dependence structure inherent in our considered class of models, stemming from skew-elliptical distributions, facilitates easier interpretation of the dependence parameters. In our analysis, we found that the use of skew-$t$ copula-based Gamma mixed models provides the best fit among eight candidate models considered, demonstrating the efficacy of our approach in capturing the complex dynamics of disease progression.
\par We derived the standard errors of the parameter estimates using the corresponding asymptotic covariance matrix (Godambe information matrix) obtained from IFM estimation. Our simulation study provides insights into the performance of IFM estimation of model parameters across various choices of multivariate copulas and marginal distributions. However, it's worth noting that even with IFM estimation, skew-$t$ and skew-normal copula-based models require significantly more computation time compared to Student-$t$ and Gaussian copula-based models. To manage the computational complexity, we employed the Gauss-Hermite quadrature rule for numerical integrations. Nevertheless, the computational burden escalates exponentially with the dimension of the random effects. In future research, we aim to explore alternative estimation methods for the proposed class of models, seeking approaches that offer one-step estimation of the model parameters. Bayesian methods present a promising avenue in this regard, enabling us to assess the impact of the copula on the estimation of regression coefficients. Furthermore, we are eager to incorporate mechanisms for handling missing data within our models, thereby enhancing their generality and versatility. This extension would enable more robust analysis in scenarios where data may be incomplete or missing. By developing techniques to effectively address missing data, we can improve the reliability and accuracy of our model estimates, leading to more meaningful insights into the underlying processes being studied. This will be particularly valuable in real-world applications where missing data is common, ensuring that our models can accommodate diverse and complex datasets encountered in practice.
\section{Declarations}
\textbf{Important Note}: This document is again updated as single authored and will not going to be communicated into any journals in the future. The readers of this manuscript are requested to cite the version $4$ if some future progress are made on the raised issues in this research. The readers are also requested to not to revisit the earlier versions. \\ \\
\bibliographystyle{agsm}
\bibliography{ref2}
\appendix
\section{Appendix}\label{apndx1}
\textbf{Assumptions for Theorem \ref{theorem1}:}
\begin{enumerate}
    \item The support of $\mathbf{Z} = (\mathbf{Y}^\intercal,\mathbf{b}^\intercal)^\intercal$, $\mathcal{Z}$ does not depend on any $\mathbf{\theta^*} \in \mathbf{\Theta^*}$.
    \item The partial derivatives $\partial\Psi^*_m/\partial\mathbf{\theta^*}$ exist for almost every $\mathbf{z} \in \mathcal{Z}$.
    \item (a) For all $\mathbf{\theta^*} \in \mathbf{\Theta^*}$,
    \begin{align}
    \frac{1}{m}\sum_{i=1}^m\Psi_{i1m}(\mathbf{\theta}) & \overset{p}{\rightarrow} \lim_{m \rightarrow \infty} \frac{1}{m}\sum_{i=1}^m E[\Psi_{i1m}(\mathbf{\theta})] = \mathbf{0}, \nonumber \\ \frac{1}{m}\sum_{i=1}^m\Psi_{i2m}(\mathbf{\phi}) & \overset{p}{\rightarrow} \lim_{m \rightarrow \infty} \frac{1}{m}\sum_{i=1}^m E[\Psi_{i2m}(\mathbf{\phi})] = \mathbf{0}. \nonumber
    \end{align}
    (b) For all $\mathbf{\theta^*} \in \mathbf{\Theta^*}$,
    \begin{align}
    & E[\Psi_{im}(\mathbf{\theta^*})\Psi_{im}(\mathbf{\theta^*})^\intercal] \;\; \text{and} \;\; \lim_{m \rightarrow \infty} \frac{1}{m}\sum_{i=1}^m E[\Psi_{im}(\mathbf{\theta^*})\Psi_{im}(\mathbf{\theta^*})^\intercal] \;\; \text{exist}, \nonumber \\ \text{and} \;\; & \frac{1}{m}\sum_{i=1}^m \Psi_{im}(\mathbf{\theta^*})\Psi_{im}(\mathbf{\theta^*})^\intercal \overset{p}{\rightarrow} \lim_{m \rightarrow \infty} \frac{1}{m}\sum_{i=1}^m E[\Psi_{im}(\mathbf{\theta^*})\Psi_{im}(\mathbf{\theta^*})^\intercal] = M_\Psi(\mathbf{\theta^*}) \nonumber
    \end{align}
    where $M_\Psi(\mathbf{\theta^*})$ is a positive definite matrix.
    \begin{equation}
    E\Big[\frac{\partial}{\partial\mathbf{\theta^*}}\Psi_{im}(\mathbf{\theta^*})\Big] \;\; \text{exists}, \; \text{and} \;\; \frac{1}{m}\sum_{i=1}^m \frac{\partial}{\partial\mathbf{\theta^*}}\Psi_{im}(\mathbf{\theta^*}) \overset{p}{\rightarrow} \lim_{m \rightarrow \infty} \frac{1}{m}\sum_{i=1}^m E\Big[\frac{\partial}{\partial\mathbf{\theta^*}}\Psi_{im}(\mathbf{\theta^*})\Big] = D_\Psi(\mathbf{\theta^*}) \nonumber     
    \end{equation}
    where $D_\Psi(\mathbf{\theta^*})$ is a non-singular matrix.
    \item The order of integration and difference can be interchanged as follows
    \begin{equation}
    \frac{\partial}{\partial\mathbf{\theta^*}}\int_{\mathcal{Z}}f^*(z,\mathbf{\theta^*})dz = \int_{\mathcal{Z}}\frac{\partial}{\partial\mathbf{\theta^*}}f^*(z,\mathbf{\theta^*})dz. \nonumber 
    \end{equation}
    \item For all $\epsilon > 0$ and for any fixed vector $\mathbf{u},(||\mathbf{u}|| \neq 0)$, the following condition is satisfied.
    \begin{equation}
    \lim_{m \rightarrow \infty} \frac{1}{m}\sum_{i=1}^m E\Big[(\mathbf{u}^\intercal\Psi_{im}(\mathbf{\theta}^*_0))^2 I\Big\{|\mathbf{u}^\intercal\Psi_{im}(\mathbf{\theta}^*_0)| \geq \epsilon\sqrt{m}
    \Big\}\Big] = 0. \nonumber
    \end{equation}
\end{enumerate}
\textbf{Proof of Thorem \ref{theorem1}:} Using Taylor's (Lagrange) expansion to the first order, we have
\begin{align}\label{expr1}
\Psi_{1m}(\hat{\mathbf{\theta}}) & = \Psi_{1m}(\mathbf{\theta}_0) + (\hat{\mathbf{\theta}} - \mathbf{\theta}_0) \frac{\partial}{\partial\mathbf{\theta}}\Psi_{1m}(\mathbf{\theta})\Big|_{\mathbf{\theta}_1}, \nonumber \\ \Psi_{2m}(\hat{\mathbf{\phi}}) & = \Psi_{2m}(\mathbf{\phi}_0) + (\hat{\mathbf{\phi}} - \mathbf{\phi}_0) \frac{\partial}{\partial\mathbf{\phi}}\Psi_{2m}(\mathbf{\phi})\Big|_{\mathbf{\phi}_1}
\end{align}
where $\mathbf{\theta}_1$ is some vector value between $\mathbf{\theta}_0$ and $\hat{\mathbf{\theta}}$, and $\mathbf{\phi}_1$ is some vector value between $\mathbf{\phi}_0$ and $\hat{\mathbf{\phi}}$, respectively. Note that $\Psi_{2m}(\hat{\mathbf{\phi}})$ also depends on the value $\hat{\mathbf{\theta}}$. Assumption $3$(a) implies
\begin{equation}
E\Big[\frac{\partial}{\partial\mathbf{\theta}}\log \int \prod_{j=1}^{n_i} f(y_{ij}|\mathbf{b_i},\theta_{ij}) g(\mathbf{b_i}) d\mathbf{b_i}\Big] \nonumber
\end{equation}
exists and naught for all $i = 1,\dots,m$. Thus we have
\begin{equation}
\frac{1}{m}\Psi_{1m}(\mathbf{\theta}_0) \overset{p}{\rightarrow} \mathbf{0}, \;\; \frac{1}{m}\Psi_{2m}(\mathbf{\phi}_0) \overset{p}{\rightarrow} \mathbf{0}. \nonumber
\end{equation}
Also from assumption $3$(b), the expectations of
\begin{equation}
\frac{1}{m}\frac{\partial}{\partial\mathbf{\theta}}\Psi_{1m}(\mathbf{\theta}) \;\; \text{and} \;\; \frac{1}{m}\frac{\partial}{\partial\mathbf{\phi}}\Psi_{2m}(\mathbf{\phi}) \nonumber
\end{equation}
converges to non-zero real vectors almost surely. Since all terms on the right hand side converges to zero, when $\hat{\mathbf{\theta}}$ and $\hat{\mathbf{\phi}}$ are the solutions of \ref{estequ1}. Hence we must have
\begin{equation}
\hat{\mathbf{\theta}} \overset{p}{\rightarrow} \mathbf{\theta}_0 \;\; \text{and} \;\; \hat{\mathbf{\phi}} \overset{p}{\rightarrow} \mathbf{\phi}_0. \nonumber
\end{equation}
\par To derive the asymptotic normality let
\begin{equation}
H_m(\mathbf{\theta^*}) = \left(\begin{array}{cc} \frac{\partial}{\partial\mathbf{\theta}}\Psi_{1m}(\mathbf{\theta}) & \mathbf{0} \\ \\ \mathbf{0} & \frac{\partial}{\partial\mathbf{\phi}}\Psi_{2m}(\mathbf{\phi})
\end{array}\right) \;\; \text{and} \;\; H_m^1 = \left(\begin{array}{cc} \frac{\partial}{\partial\mathbf{\theta}}\Psi_{1m}(\mathbf{\theta})\Big|_{\mathbf{\theta}_1} & \mathbf{0} \\ \\ \mathbf{0} & \frac{\partial}{\partial\mathbf{\phi}}\Psi_{2m}(\mathbf{\phi})\Big|_{\mathbf{\phi}}
\end{array}\right) \nonumber
\end{equation}
We rewrite the expression in \ref{expr1} as
\begin{equation}\label{expr2}
\sqrt{m}(\hat{\mathbf{\theta}}^* - \mathbf{\theta}^*_0) = \Big[\frac{1}{m}H_m^1\Big]^{-1}\frac{1}{\sqrt{m}}[-\Psi_m(\mathbf{\theta}^*_0)].
\end{equation}
Since $\hat{\mathbf{\theta}}^*$ is a consistent estimator of $\mathbf{\theta}^*_0$, from the convergence in probability we have
\begin{equation}
\frac{1}{m}[H_m(\hat{\mathbf{\theta}}^*) - H_m(\mathbf{\theta}^*_0)] \overset{p}{\rightarrow} \mathbf{0}. \nonumber     
\end{equation}
Now from assumption $3$(b) we have,
\begin{equation}
\frac{1}{m}H_m(\mathbf{\theta}^*_0) = \left(\begin{array}{cc} \frac{1}{m}\sum_{i=1}^m \frac{\partial}{\partial\mathbf{\theta}}\Psi_{i1m}(\mathbf{\theta})\Big|_{\mathbf{\theta}_0} & \mathbf{0} \\ \\ \mathbf{0} & \frac{1}{m}\sum_{i=1}^m \frac{\partial}{\partial\mathbf{\phi}}\Psi_{i2m}(\mathbf{\phi})\Big|_{\mathbf{\phi}}
\end{array}\right) \;\; \overset{p}{\rightarrow} D_\Psi(\mathbf{\theta}^*_0).
\nonumber
\end{equation}
Thereafter using assumption $3$(b) and $4$ we get,
\begin{align}
\frac{1}{m^2} Cov[H_m(\mathbf{\theta}^*_0)] & = \frac{1}{m^2}\sum_{i=1}^m Cov\Big[\frac{\partial}{\partial\mathbf{\theta^*}}\Psi_{im}(\mathbf{\theta^*})\Big|_{\mathbf{\theta}^*_0} \nonumber \\ & = \frac{1}{m}\Big[\frac{1}{m}\sum_{i=1}^m E\Big[\frac{\partial}{\partial\mathbf{\theta^*}}\Psi_{im}(\mathbf{\theta^*})\Big|_{\mathbf{\theta}^*_0}\frac{\partial}{\partial\mathbf{\theta^*}^\intercal}\Psi_{im}(\mathbf{\theta^*})^\intercal\Big|_{\mathbf{\theta}^*_0}\Big] \nonumber \\ & = \frac{1}{m}\Big[\frac{1}{m}\sum_{i=1}^m \frac{\partial^2}{\partial\mathbf{\theta^*}\partial\mathbf{\theta^*}^\intercal} E\Big[\Psi_{im}(\mathbf{\theta^*})\Big|_{\mathbf{\theta}^*_0}\Psi_{im}(\mathbf{\theta^*})^\intercal\Big|_{\mathbf{\theta}^*_0}\Big] \;\; \rightarrow \mathbf{0} \;\; \text{as} \;\; m \rightarrow \infty. \nonumber
\end{align}
Now $\mathbf{\theta}^*_1$ lies in between $\hat{\mathbf{\theta}}^*$ and $\mathbf{\theta}^*_0$, thus by weak law of large number
\begin{equation}
\frac{1}{m} H_m^1 - D_\Psi(\mathbf{\theta}^*_0) \;\; \overset{p}{\rightarrow} \mathbf{0}. \nonumber
\end{equation}
The final term of the expression \ref{expr2}, $\Psi_m(\mathbf{\theta}^*_0)$ involves sum of independent terms, which have expectation $\mathbf{0}$ and covariance matrix $Cov[\Psi_{im}(\mathbf{\theta}^*_0)] = E[\Psi_{im}(\mathbf{\theta}^*_0)\Psi_{im}(\mathbf{\theta}^*_0)^\intercal]$ for $i = 1,\dots,m$. Hence from assumption $5$, with direct application of Lindeberg-Feller central limit theorem, for any fixed vector $\mathbf{u}$ we have,
\begin{equation}
\mathbf{u}^\intercal\Big(\frac{\Psi_m(\mathbf{\theta}^*_0)}{\sqrt{m}}\Big) \;\; \overset{d}{\rightarrow} N(0,\mathbf{u}^\intercal M_\Psi(\mathbf{\theta}^*_0)\mathbf{u}) \nonumber
\end{equation}
Combining everything and using Slutsky's theorem we finally have,
\begin{equation}
\sqrt{m}(\hat{\mathbf{\theta}}^* - \mathbf{\theta}^*_0) \;\; \overset{d}{\rightarrow} N(\mathbf{0},J_\Psi(\mathbf{\theta}^*_0)^{-1}), \;\; \text{where} \;\; J_\Psi(\mathbf{\theta}^*_0) = D_\Psi(\mathbf{\theta}^*_0)^\intercal M_\Psi(\mathbf{\theta}^*_0)^{-1} D_\Psi(\mathbf{\theta}^*_0), \nonumber
\end{equation}
and that completes the proof.
\end{document}